
\input phyzzx
\input tables \date={1992}
\rightline {November 1992} \rightline {SUSX--TH--92/17.}
\title {Twisted Sector Yukawa Couplings For The ${\bf Z}_M\times
{\bf Z}_N$ Orbifolds}
\author{D. Bailin${a}$, \ A. Love${b}$ \ and \ W.A. Sabra${c}$\foot{Address
from January 1993, Department of Physics, Royal Holloway and Bedford New
College, University of London, Surrey, U.K.}}
\address {${a}$School of Mathematical and Physical
Sciences,\break
University of Sussex, \break Brighton U.K.}
\address {${b}$Department of Physics,\break
Royal Holloway and Bedford New College,\break
University of London,\break
Egham, Surrey, U.K.}
\address {${c}$Physics Department \break Birkbeck College,\break
University of London,\break
Malet Street\break London WC1E 7HX}
\abstract{The moduli dependent Yukawa couplings between twisted
sectors of \hfill\break
${\bf Z}_M\times {\bf Z}_N$ Coxeter orbifolds are
studied.}
\endpage
\REF\one{L. Dixon, J. A. Harvey, C. Vafa and E. Witten, Nucl. Phys.
B261 (1985) 678; B274 (1986) 285.}
\REF\two{ A. Font, L. E. Ibanez,
F. Quevedo and A. Sierra, Nucl. Phys. B331 (1991) 421.}
\REF\three{L. Dixon, D. Friedan, E. Martinec and S. Shenker, Nucl.
Phys. B282 (1987) 13.}
\REF\four{S. Hamidi, and C. Vafa, Nucl. Phys. B279 (1987) 465.}
\REF\five{ L. E. Ibanez, Phys. Lett. B181 (1986) 269.}
\REF\six{J. A. Casas and C. Munoz, Nucl. Phys. B332 (1990) 189}
\REF\seven {J. A. Casas, F. Gomez and C. Munoz,  Phys. Lett.
B251(1990) 99}
\REF\eight {J. A. Casas, F. Gomez and C. Munoz,  CERN preprint,
TH6194/91.}
\REF\nine{T. Kobayashi and N. Ohtsubu, Kanazawa
preprint, DPKU--9103.}
\REF\ten {T. T. Burwick, R. K. Kaiser and H. F.
Muller, Nucl. Phys. B355 (1991) 689}
\REF\eleven{J. Erler, D. Jungnickel and J. Lauer, Phys. Rev D45
(1992) 3651; S. Stieberger, D. Jungnickel, J. Lauer
and M. Spalinski, preprint, MPI--Ph/92--24; J. Erler,  D.
Jungnickel, M. Spalinski and S. Stieberger, preprint,
MPI--Ph/92--56.}
\REF\twelve{D. Bailin, A. Love and W. A. Sabra,
Mod. Phys. Lett A6 (1992) 3607.}
\REF\thirteen{S. Stieberger, preprint, TUM--TH--151/92.}
\REF\fourteen{V. S. Kaplunovsky, Nucl. Phys. B307 (1988) 145; L.
J. Dixon, V. S. Kaplunovsky and J. Louis,  Nucl. Phys. B355 (1991)
649; J. P. Derenddinger, S. Ferrara, C. Kounas and F. Zwirner,
Nucl. Phys. B372 (1992) 145, Phys. Lett. B271 (1991) 307.}
\REF\fifteen{I. Antoniadis, J. Ellis, R. Lacaze and D. V. Nanopoulos,
CERN--TH \hfill\break
6136/91; S. Kalara, J. L. Lopez and D. V. Nanopoulos,
Phys. Lett. B269 (1991) 84.}
\REF\sixteen {L. E. Ibanez and D. L\"ust
and G. G. Ross, Phys. Lett. B272 (1991) 251; D. Bailin and A. Love,
Phys. Lett. B278 (1992) 125.}
\REF\seventeen{L. E. Ibanez and D.
L\"ust, CERN--TH 6380/92.}
\REF\eighteen{D. Bailin and A. Love,
Phys. Lett. B292 (1992) 315.}
\REF\nineteen{Y.
Katsuki, Y. Kawanura,  T. Kobayashi and N. Ohtsubo,Y. Ono and K.
Tanioka, Nucl. Phys. B341 (1990) 611; D. Narkushevich, N.
Olshanetsky and A. Perelonov, Comm. Math. Phys. 111 (1987) 247.}
\REF\twenty{T. Kobayashi and N. Ohtsubo, Phys. Lett. B262 (1991)
425.}
\REF\twentyone{D. J. Gross, J. A. Harvey, E. Martinec and R.
Rohm, Nucl. Phys. B267 (1986) 75.}
\REF\twentytwo{A. Font, L. E.
Ibanez and F. Quevedo, Phys. Lett. B217 (1989) 272.}
\REF\twentythree{T. Kobayashi and N. Ohtsubo, Phys. Lett. B245
(1990) 441.}
\chapter{Introduction}
A knowledge of the Yukawa couplings for orbifold compactified
string theory models [\one, \two] is an important step towards
comparison of the models with observation. Of particular importance
is the exponential dependence of Yukawa couplings on moduli
[\three, \four] which can arise when all the states involved are in
twisted sectors, because of its possible bearing on hierarchies
[\five]. Twisted sector Yukawa coulping have been investigated for
the ${\bf Z}_3$ orbifold [\three--\six], for the ${\bf Z}_7$ orbifold
[\seven] and for arbitrary ${\bf Z}_N$ orbifolds [\eight--\eleven].
Here, we shall extend our recent discussion [\twelve] of Yukawa
couplings for the ${\bf Z}_3\times {\bf Z}_3$ orbifold to all
Coxeter ${\bf Z}_M\times {\bf Z}_N$ orbifolds with the point group
realized by Coxeter elements acting on a direct sum of
two-dimensional root lattices. After completion of the present work
we have received a preprint by Stieberger [\thirteen] which also
addresses the question of Yukawa couplings for ${\bf Z}_M\times {\bf
Z}_N$ orbifolds though with the
emphasis on general properties such as modular symmetries rather
than specific orbifolds on specific lattices. A possible advantage
that ${\bf Z}_M\times {\bf Z}_N$ orbifolds possess over ${\bf Z}_N$
orbifolds is that they can more generically provide the string loop
threshold corrections needed to explain the low energy values of the
gauge coupling constants [\fourteen--\eighteen] because they are not
so tightly constrained by duality anomaly cancellation conditions
[\seventeen]. \chapter{Yukawa couplings for ${\bf Z}_M\times {\bf
Z}_N$ Orbifolds.} To specify an orbifold model it is necessary to
specify both the point group and the lattice on which the point
group acts as a symmetry. Perhaps the most elegant possibility
[\nineteen] is that the point group acts as a Coxeter element
(product of Weyl reflections) on a root lattice of a Lie algebra.
Here, we shall study the case where the root lattice involved is a
direct sum of three 2-dimensional root lattices. For the ${\bf
Z}_M\times {\bf Z}_N$ orbifolds (table 1) the factors involved are
${\bf Z}_3$, ${\bf Z}_4$, ${\bf Z}_6$ and ${\bf Z}_2$. These
discrete groups can be realised as the Coxeter elements for the
$SU(3)$, $SO(5)$, $G_2$ and $SO(4)$ root lattices, respectively
[\twenty, \nineteen]. The ${\bf Z}_2$ group can also be realized as
the square of the Coxeter element of the $SO(5)$ lattice and the
cube of the Coxeter element on the $G_2$ lattice, and the ${\bf
Z}_3$ group can also be realized as the square of the Coxeter
element on the $G_2$ lattice. The simplest possibility (table 1 of
the present work, and table 1 of ref. 20) is to realize  ${\bf
Z}_3$, ${\bf Z}_4$ and ${\bf Z}_6$ as a Coxeter element on each of
the two dimensional sub-lattices (rather than as a power of Coxeter
element). This can be done for each of the  ${\bf Z}_M\times {\bf
Z}_N$ orbifolds, except ${\bf Z}_3\times {\bf Z}_6$, by the choice
of lattices in table 1, and for ${\bf Z}_3\times {\bf Z}_6$ it can
be done for 2 of the 3 sub-lattices with ${\bf Z}_3$ realized as the
square of the Coxeter element on the third $G_2$ sub-lattice. These
are  the ${\bf Z}_M\times {\bf Z}_N$ models we shall study here.

The action of the Coxeter element $C(G)$ of a rank 2 Lie algebra
$G$ on the basis vectors ${e}_1$ and ${e}_2$ of the root
lattices may be taken to be
$$C(SU(3))e_1=e_2,\quad C(SU(3))
e_2=-e_1-e_2,\eqn\le$$ $$C(SO(5))e_1=e_1+2
e_2,\quad C(SO(5))e_2=-e_1-e_2,\eqn\lebb$$
$$C(G_2)e_1=-e_1-e_2,\quad C(G_2)e_2=3e_1+2e_2,\eqn\lebba$$ and
$$C(SO(4))e_1=-e_1,\quad C(SO(4))e_2=-e_2\eqn\lebban$$
realizing ${\bf Z}_3$, ${\bf Z}_4$, ${\bf Z}_6$ and ${\bf Z}_2$,
respectively.  The powers of Coxeter elements $C2(SO(5))$ and
$C3(G_2)$ realizing ${\bf Z}_2$ have the same action as $C(SO(4))$,
and $C2(G_2)$ realizing ${\bf Z}_3$ has the action
$$C2(G_2)e_1=-2e_1-e_2,\quad
C2(G_2)e_2=3e_1+e_2.\eqn\lebbanon$$

The non-zero Yukawa couplings obey point group selection rules, and
selection rules for the H-lattice momentum associated with
bosonized right-moving NSR fermionic degrees of freedom. These
selection rules have already been written down [\twenty] for
${\bf Z}_M\times {\bf Z}_N$ orbifolds. They also obey space group selection
rules [\three,\four] which may be written in the following form.
Let $(\alpha, l_1)$, $(\beta, l_2)$ and $(\gamma, l_3)$ be space
group elements associated with the $\alpha$, $\beta$ and
$\gamma$ twisted sectors, where
$$l_1=(I-\alpha)f_\alpha+(I-\alpha)\Lambda,\eqn\n$$
$$l_2=(I-\beta)f_\beta+(I-\beta)\Lambda,\eqn\m$$
and
$$l_3=(I-\gamma)f_\gamma+(I-\gamma)\Lambda.\eqn\p$$
In \m--\p, $f_\alpha$, $f_\beta$ and $f_\gamma$ are fixed points or
fixed tori for the $\alpha$, $\beta$ and
$\gamma$ twisted sectors, respectively, and in each case $\Lambda$
denotes an arbitrary lattice vector. Then, given that we have
already satisfied the point group selection rule
$$\alpha\beta\gamma=I,\eqn\po$$
the full space group selection rule contains the additional
condition
$$l_1 +l_2+ l_3\ \hbox{contains}\ 0.\eqn\sel$$
The values of the non-zero Yukawa couplings amongst the twisted
sectors are determined in detail by three--point functions involving
fermionic and bosonic degrees of freedom. However, the crucial
dependence on the deformation parameters (moduli) and the
particular fixed points and fixed tori
is entirely contained in (bosonic) twist field correlation
functions [\three,\four] of the type
$${\cal Z}=\prod_{i=1}3<\sigma_\alphai(z_1, \bar z_1)
\sigma_\betai(z_2, \bar z_2)\sigma_\gammai(z_3,
\bar z_3)>=\prod_{i=1}3 {\cal Z}_i,\eqn\cor$$ where
$\alpha,$ $\beta$ and
$\gamma$ now label the twisted sectors and also the fixed
points or fixed tori involved, and the index $i$ distinguishes the
twist fields associated with the complex coordinates $X_i$,
$i=1,2,3,$ that define the six-dimensional compact manifold. The
correlation function
$${\cal Z}_i=<\sigma_\alphai(z_1, \bar z_1)
\sigma_{\beta}i(z_2, \bar z_2)\sigma_{\gamma}i(z_3,
\bar z_3)>\eqn\leb$$
factors [\three, \four] into a quantum piece ${\cal Z}i_{qu}$ and a
classical piece with all the dependence on the moduli and the
particular fixed points and fixed tori involved contained in the
classical piece.
$${\cal Z}_i={\cal Z}i_{qu} \sum_{X_{cl}}e{-Si_{cl}},\eqn\leba$$
where the classical action is
$$Si_{cl}={1\over\displaystyle \pi}\int d2z\Big
({\partial X_i\over\displaystyle\partial
z}{\partial \bar X_i\over\displaystyle  \partial \bar z}+{\partial
 X_i\over\displaystyle
\partial \bar z}{\partial \bar X_i\over\displaystyle  \partial
 z}\Big).\eqn\leban$$
Because of the string equations of motion
$${\partial2 X_i\over\displaystyle  \partial z\partial\bar z}=0,\eqn\lebano$$
${\partial X_i/\partial z}$ and ${\partial X_i/ \partial
\bar z}$ are functions of $z$ and $\bar z$ alone, respectively,
which have to be chosen to respect the boundary conditions at
$z_1$, $z_2$ and $z_3$ implicit in the operator product
expansions with the twist fields [\three].
Let the space group elements for the fixed points or fixed tori
involved in the Yukawa coupling be $(\alpha, l_1),$ $(\beta,
l_2)$ and $(\gamma, l_3)$, as before, with $l_1,$ $l_2$ and $l_3$
as in \n--\p. Also let $P$ be the least common denominator for the
fractional twists involved so that the action of the point group
elements $\alpha$, $\beta$ and $\gamma$ on the three complex
coordinates $X_i$ may be written in the form
$$\eqalign{&\alpha: X_i\rightarrow e{2\pi iA_i/P}X_i,\cr &
\beta: X_i\rightarrow e{2\pi iB_i/P}X_i,\cr &
\gamma: X_i\rightarrow e{2\pi iC_i/P}X_i,}\eqn\sand$$
where $A_i$, $B_i$ and $C_i$ are positive integers smaller than
$P$ with
$$A_i+B_i+C_i=0\ (\hbox{mod}\ P),\eqn\sandr$$
for consistency with the point group selection rule. When one of
the $A_i$, $B_i$ or $C_i$ is zero, that is one of the
$\alpha$, $\beta$ or $\gamma$ leaves the $i$-th complex plane
invariant, then the three twist field correlation function reduces to
a 2 twist field correlation function, which may be normalized to one.
Otherwise, we need to evaluate the classical action using
$${\partial X_i\over\displaystyle  \partial z}= a_i(z-z_1){-(1-A_i/
P)}(z-z_2){-(1-B_i/P)}(z- z_3){-(1-C_i/P)}\eqn\brit$$ and
$${\partial X_i\over\displaystyle  \partial\bar z}= b_i(\bar z-\bar
z_1){-A_i/P}(\bar z-\bar z_2){-B_i/P}(\bar z- \bar
z_3){-C_i/P}.\eqn\brita$$  In all the cases we shall study here,
only the holomorphic field ${\partial X_i/\partial z}$ is an
acceptable  classical solution, because  ${\partial X_i/\partial\bar
z}$ gives a divergent contribution to the classical action and we
must set $$b_i=0.\eqn\britan$$ The constants $a_i$ may be evaluated
by the use of global monodromy conditions [\three] where we
integrate around a closed contour ${\cal C}_i$ such that $X_i$ is
shifted by $v_i$ but not rotated. Thus, $$\oint_{{\cal C}_i} dz
{\partial X_i\over\displaystyle \partial z}=v_i,\eqn\britian$$ where we have
dropped the contribution of  ${\partial X_i/\partial\bar z}$ as just
discussed. If, for example, we choose ${\cal C}_i$ to go $B_i$ times
around $z_1$ counter-clockwise and $A_i$ times around $z_2$
clockwise, with the choice (allowed by $SL(2,C)$ symmetry)
$$z_1=0,
\quad z_2=1,\quad z_3=\infty,\eqn\eng$$ then $X_i$ is not
rotated. Multiplying together the relevant space group elements,
$(\alpha, l_1){B_i}(\beta{-1}, l_2){A_i}$ we find the shift in
$i$-th complex plane to be the projection into this plane of
$$W_i=(1-\alpha{B_i})(f_\alpha-f_\beta+\Lambda),\eqn\engl$$ where
$\Lambda$ is an arbitrary lattice vector. In the first instance,
\engl\ is in the lattice basis for the Coxeter orbifold. If we
transform to the orthonormal basis in which the action of $\alpha$,
$\beta$ and $\gamma$ is given by \sand, then $v_i$ is the component
of $W_i$ in the $i$-th complex direction, e.g. for $X_1$ the shift
$v_1$ is the component of $W_1$ in the $1+i2$ direction, where $1$
and $2$ denote the orthonormal basis. The global monodromy condition
may be evaluated with the aid of the  integral $$\oint_{{\cal C}_i}
dz z{-{(1-A_iy)}}(z-1){-{(1-B_iy)}}=-2i\sin (A_iB_i\pi
y){\Gamma(A_iy)\Gamma(B_iy)\over\displaystyle \Gamma(A_iy+B_iy)}\eqn\engla$$
 with
the result
$$2ia_i\sin({A_iB_i\pi\over\displaystyle
P}){\strut\Gamma({\displaystyle A_i\over\displaystyle \displaystyle
P})\Gamma({\displaystyle B_i\over\displaystyle  \displaystyle
P})\over\displaystyle \displaystyle \Gamma({A_i\over\displaystyle  P}+
 {B_i\over\displaystyle
P})}(-z_\infty){-(1-C_i/P)}=v_i\eqn\englan$$  which determines
$a_i$. The contribution to the classical action may now be obtained
by using Appendix A of [\twentyone], with the result
$$S_{cl}i={{\vert v_i\vert}2\over\displaystyle
4\pi\sin2({\displaystyle A_i\displaystyle
B_i\pi\over\displaystyle \displaystyle P})}{\sin({\displaystyle
A_i\pi\over\displaystyle \displaystyle P})\sin({\displaystyle
B_i\pi\over\displaystyle \displaystyle P})\over\displaystyle
 \sin({(\displaystyle
A_i+\displaystyle B_i)\pi\over\displaystyle \displaystyle P})}.\eqn\england$$

Exactly similar expressions may be written down when the
contour is chosen instead to encircle the pair of twist fields
associated with $(\alpha, l_1)$ and $(\gamma, l_3)$, or the pair of
twist fields associated with $(\beta, l_2)$ and $(\gamma, l_3).$
In general, to obtain consistency between these different global
monodromy conditions it is necessary in \leba\ to restrict the sum
over the initially arbitrary lattice vectors arising in $v_i$ in a
way which depends on the fixed tori involved. This problem has been
solved for the case of ${\bf Z}_N$ orbifolds in appendix B of the last
reference of [11], and this same result can be applied here
provided we restrict attention to one particular complex coordinate
$X_i$ at a time.

In general, after general GSO projections inclusive of Wilson lines
[\twentytwo, \two] the surviving physical states are eigenstates of
the two point group elements $\theta$ and $\omega$ generating the
${\bf Z}_M\times {\bf Z}_N$ point group. In the construction of these $\theta$
and $\omega$ eigenstates account must be taken of the fact that
fixed points (or tori) are not always left invariant [\twentythree,
\nine] by the action of $\theta$ or $\omega$. If $m$ and $n$ are
the least integers such that for the fixed point (or torus)
$f_{kl}$ of $\thetak\omegal,$
$$\thetam f_{kl}\sim f_{kl}\eqn\ra$$and
$$\omegan f_{kl}\sim f_{kl},\eqn\rag$$
where $\sim$ signifies up to a lattice vector, then the $\theta$
and $\omega$ eigenstate combinations of fixed points $\vert p, q>$
are of the form
$$\vert p, q>=\sum_{r=0}{m-1}\sum_{s=0}{n-1} e{-{i2\pi
pr/m}}e{-{i2\pi qs/n}}\vert \thetar\omegas f_{kl}>,\quad
p=0,..,m-1\quad q=0,..,n-1,\eqn\su$$
which are simultaneous eigenstates of $\theta$ and $\omega$ with
eigenvalues
$e{{i2\pi p/m}}$ and $e{{i2\pi q/n}}$, respectively.

Then, the Yukawa couplings for physical states are obtained as
combinations of terms for definite fixed tori, subject to the
space group selection rules.
\chapter{The ${\bf Z}_2\times {\bf Z}_6$ orbifold}
In this section, we shall illustrate the calculations and results
by considering the case of the ${\bf Z}_2\times {\bf Z}_6$ Coxeter orbifold on
the $SO(4)\times G_22$ lattice as in table 1. The corresponding
results for the other ${\bf Z}_M\times {\bf Z}_N$ orbifolds are given in
summary in the appendices. In the orthonormal basis, the action of
the generators $\theta$ and $\omega$ of the point group on the
complex string degrees of freedom $X_i$, $i=1,2,3,$ is
$$\theta X_1=-X_1,\quad \theta X_2=X_2,\quad \theta
X_3=-X_3,\eqn\po$$ and
$$\omega X_1=X_1,\quad \omega X_2=e{2\pi i/6}X_2,\quad
\omega X_3=e{-{2\pi i/6}}X_3.\eqn\pol$$
On the other hand, for this Coxeter orbifold, the action of
$\theta$ and $\omega$ on the lattice basis is given by
$$\theta=(C(SO(4)), I, C{-3}(G_2))\eqn\poly$$
and
$$\omega=(I, C(G_2), C{-1}(G_2)),\eqn\polya$$
where $C(H)$ is the Coxeter element for the Lie algebra of $H.$
Explicitly, using \lebba\ and \lebban, the action on the 6 basis
vectors $e_1$,...,$e_6$ for the lattice basis is
$$\eqalign{&\theta e_1=-e_1,\cr &
\theta e_2=-e_2},\qquad
\eqalign{&\theta e_3=e_3,\cr &\theta e_4=
e_4},\qquad \eqalign{&\theta e_5=-e_5,\cr &\theta
e_6=-e_6}\eqn\polyak$$  and $$\eqalign{&\omega e_1=e_1,\cr &\omega
e_2=e_2},\qquad\eqalign{&\omega e_3=-e_3-
e_4,\cr &\omega e_4=3e_3+2e_4},\qquad\eqalign{&\omega e_5=2
e_5+e_6,\cr &\omega e_6=-3e_5-e_6.}\eqn\polyako$$  More generally,
the original rigid lattice may be deformed in ways which preserve
the action \polyak\ and \polyako\ of the point group. To obtain the
most general choice of lattice compatible with the point group we
must require that all the scalar products $e_i.e_j$ are
preserved by the transformations \polyak\ and \polyako. If we write
$$e_i.e_j=\vert e_i\vert\vert
e_j\vert\cos\theta_{ij},\quad i\not=j,\eqn\polyakov$$ we find that
$$\vert e_4\vert=\sqrt 3\vert e_3\vert\equiv \sqrt 3R_3,\eqn\wi$$
$$\vert e_6\vert=\sqrt 3\vert e_5\vert\equiv \sqrt 3R_5,\eqn\wit$$
and  $$\cos\theta_{34}=\cos\theta_{56}=-{\sqrt 3\over\displaystyle
2},\eqn\witt$$ with all other $\cos\theta_{ij}$ zero except
$\cos\theta_{12},$ and $$\vert e_1\vert\equiv R_1,\eqn\witte$$
$$\vert e_2\vert\equiv R_2,\eqn\witten$$ and $\cos\theta_{12}$
undetermined.  Thus we may take 5 independent deformations of the
lattice compatible with the point group to be $R_1$, $R_2$, $R_3$,
$R_5$ and $\cos\theta_{12}$. These are the deformation parameters or
moduli. The fixed points and fixed tori for the twisted sectors are
readily obtained using the action \polyak\ and \polyako\ of the
generators of the point group on the (deformed) lattice. These are
presented in table 2, for the twisted sectors containing massless
states together with the associated space group elements. The space
group selection rules for the various Yukawa couplings amongst
twisted sectors are then easily obtained from \sel. All Yukawa
couplings consistent with the point group selection rule and
H-momentum conservation rule for the bosonized NSR fermion right
movers have already been listed in [\twenty].
As already discussed in section 2, the physical states surviving
generalized GSO projections are $\theta$ and $\omega$ eigenstates,
which particular eigenstates survive depending on the details of
the point group embeddings and Wilson lines in a particular model.
The $\theta$ and $\omega$ eigenstate combinations of fixed points
(or tori) may be obtained from the action of $\theta$ and $\omega$
on the fixed tori given in table 3. An orthonormal basis $\tilde
e_1,...,\tilde e_6$ in which $X_1$, $X_2$ and $X_3$ are
the components of the string degrees of freedom in the $1+i2$, $2+i3$
and $5+i6$ directions, respectively, is obtained by comparing
\po\ and \pol\ with \polyak\ and \polyako, and using \wi--\witt\
when ensuring the orthonormality of the basis. A suitable choice is
related to the lattice basis by $$\eqalign{&e_1=R_1\tilde
e_1,\cr & e_2=R_2(\cos\theta_{12}\tilde e_1+\sin\theta_{12}\tilde
e_2),}\quad \eqalign{& e_3=R_3\tilde e_3,\cr &
e_4=-R_3({3\over\displaystyle 2}\tilde e_3+{\sqrt
3\over\displaystyle 2}\tilde e_4),}\quad \eqalign{& e_5=R_5\tilde
e_5,\cr &  e_6=-R_5({3\over\displaystyle 2}\tilde e_5+{\sqrt
3\over\displaystyle  2}\tilde e_6).}\eqn\al$$

The moduli and fixed point (or tori) dependence of the non-zero
Yukawa couplings for twisted sectors containing massless states may
now be calculated using the approach described in section 2. Four
types of contribution to the Yukawa coupling arise  for the
${\bf Z}_2\times {\bf Z}_6$ orbifold and also for the
${\bf Z}_3\times {\bf Z}_6,$ ${\bf Z}_2\times {\bf Z}'_6$ and ${\bf
Z}_6\times {\bf Z}_6$ orbifolds. First, if ${\cal Z}_i$ of \leb\
reduces to a two point function because the $i$-th complex plane is
unrotated in one of the three twisted sectors then we may normalize
${\cal Z}_i$ to 1. On the other hand, if all the three twist fields
in \leb\ are non-trivial then the contribution to $Si_{cl}$ takes
one of three forms.  When all three twists are $e{2\pi i/3}$, then
$$Si_{cl}={R2_{2i-1}\over\displaystyle  8\pi\sqrt
3}\Big[(2m_{2i-1}-2n_{2i-1}+3k_{2i-1}+6k_{2i})2+3k_{2i-1}2\Big]
\eqn\ale$$
where $i=1, 2 $ or 3, the arbitrary integers $k_{2i-1}$ and $k_{2i}$
derive from the freedom in the choice of the space group elements
associated with particular fixed tori, and the integers
$m_{2i-1}$ and $n_{2i-1}$ characterize any two of the fixed tori
involved. The leading contribution to the Yukawa coupling in
\leba\ is obtained by
minimizing $Si_{cl}$ by varying the arbitrary integers $k_{2i-1}$
and $k_{2i}$. In the case of \ale, the minimum value of
$Si_{cl}$ is given by
$$\eqalign{\Big(S_{cl}i\Big)_{min}=&0,\qquad m_{2i-1}-n_{2i-1}=0\cr
=&{2R_{2i-1}2\over\displaystyle  4\pi \sqrt 3},\qquad
m_{2i-1}-n_{2i-1}=\pm 1,\pm 2.}\eqn\alex$$
The corresponding moduli dependent suppression factor in the Yukawa
coupling is $\exp{(-Si_{cl})_{min}}.$

For one twist of $e{4\pi i/3}$ and two twists of
$e{2\pi i/6},$
$$Si_{cl}={R2_{2i-1}\over\displaystyle  16\pi\sqrt
3}\Big[(2n_{2i-1}+3k_{2i-1}+6k_{2i})2+3k_{2i-1}2\Big]
\eqn\ba$$
where $n_{2i-1}$ characterizes the fixed torus in the sector where
the twist on $X_i$ is $e{4\pi i/3}.$ In this
case, the minimum value of $Si_{cl}$ is given by
$$\eqalign{\Big(S_{cl}i\Big)_{min}=&0,\qquad n_{2i-1}=0\cr
=&{R_{2i-1}2\over\displaystyle  4\pi \sqrt 3},\qquad n_{2i-1}=\pm
1.}\eqn\bai$$
Finally, for twists of $e{2\pi i/6},$ $e{2\pi i/3}$
and $-1$,
$$Si_{cl}={R2_{2i-1}\over\displaystyle  16\pi\sqrt
3}\Big[(2n_{2i-1}+3p_{2i}+6k_{2i})2+3(2n_{2i-1}+3p_{2i}-2p_{2i-1}+
6k_{2i}+4k_{2i-1})2\Big],\eqn\baili$$
where $n_{2i-1}$ characterizes the fixed torus in the sector where
the twist on $X_i$ is $e{2\pi i/3},$ and
$p_{2i-1}$ and $p_{2i}$ characterize the fixed torus in the sector
where the twist is $-1.$ In this case, the minimum value of
$Si_{cl}$ is given by $$\eqalign{\Big(S_{cl}i\Big)_{min}=&0,\qquad
n_{2i-1}=p_{2i-1}=p_{2i}=0\cr =&{3R_{2i-1}2\over\displaystyle  4\pi
\sqrt 3},\qquad n_{2i-1}=0,\ p_{2i-1}, p_{2i}\ \hbox{not\ both\
zero}\cr =& {4R_{2i-1}2\over\displaystyle  4\pi \sqrt 3},\qquad
n_{2i-1}=\pm 1,\ p_{2i-1}=p_{2i}=0 \cr =&
{R_{2i-1}2\over\displaystyle  4\pi \sqrt 3},\qquad
n_{2i-1}=\pm 1,\ p_{2i-1}, p_{2i}\ \hbox{not\ both\ zero.}}
\eqn\bailin$$

The suppression factors for the non-zero Yukawa couplings
amongst the twisted sectors of the ${\bf Z}_2\times {\bf Z}_6$ orbifold are
tabulated in table 4. Corresponding results for the other
${\bf Z}_M\times {\bf Z}_N$ orbifolds are given in the appendix. For the
${\bf Z}_2\times {\bf Z}_4$ and ${\bf Z}_4\times {\bf Z}_4$ orbifolds, the only
 non-trivial
suppression factor arises when the twists on $X_i$ are $e{2\pi
i/4}$, $e{2\pi i/4}$ and $-1$. Then,
$$Si_{cl}={R2_{2i-1}\over\displaystyle16\pi
}\Big[(2n_{2i}+2p_{2i-1}-p_{2i}+4k_{2i-1}+2k_{2i})2+
(p_{2i}-2k_{2i})2\Big],
\eqn\hellraiser$$
where $n_{2i}$ characterizes the fixed torus in one of the sectors
where the twist on $X_i$ is $e{2\pi i/4}$ and $p_{2i-1}$
and $p_{2i}$ characterize the fixed torus in the sector
where the twist is $-1.$ In this case, the minimum value of
$Si_{cl}$ is given by
$$\eqalign{\Big(S_{cl}i\Big)_{min}=&{R_{2i-1}2\over\displaystyle
8\pi},\qquad
p_{2i}=1\cr =&0,\qquad p_{2i}=0,\ n_{2i}+p_{2i-1}=0\
(\hbox{mod}\ 2)\cr =& {R_{2i-1}2\over\displaystyle  4\pi},\qquad
p_{2i}=0,\ n_{2i}+p_{2i-1}=1\
(\hbox{mod}\ 2).}\eqn\pinhead$$
For all cases other than the ${\bf Z}_2\times
{\bf Z}_4$ and ${\bf Z}_4\times {\bf Z}_4$  orbifolds, the suppression factors
 are
written in the form  $e{-(\lambda_{2i-1}R_{2i-1}2/4\pi\sqrt 3)}$,
and for the  ${\bf Z}_2\times {\bf Z}_4$ and ${\bf Z}_4\times {\bf Z}_4$
 orbifolds in the
form  $e{-(\lambda_{2i-1}R_{2i-1}2/16\pi)}.$
The suppression factors arising from \bailin\ are summarized by
$$\lambda_i=\mu(n_{2i-1}, p_{2i-1}, p_{2i}),\eqn\bosnia$$
where
$$\eqalign{\mu(n_{2i-1}, p_{2i-1}
p_{2i})=&0,\qquad
n_{2i-1}=p_{2i-1}=p_{2i}=0\cr =&3,\qquad
n_{2i-1}=0,\ p_{2i-1}, p_{2i}\ \hbox{not\ both\
zero}\cr =&4, \qquad n_{2i-1}=\pm 1,\ p_{2i-1}=p_{2i}=0\cr =&
1,\qquad n_{2i-1}=\pm 1,\  p_{2i-1}, p_{2i}\ \hbox{not\ both\
zero,}}\eqn\somalia$$ where $n_{2i-1}$
characterizes the fixed torus in the sector
where the twist on $X_i$ is $e{2\pi i/3},$ and
$p_{2i-1}$ and $p_{2i}$ characterize the fixed torus in the sector
where the twist is $-1.$
The suppression factors arising from \pinhead\ are summarized by
$$\lambda_i=\rho(n_{2i}, p_{2i-1}, p_{2i}),\eqn\tv$$
where
$$\eqalign{\rho(n_{2i}, p_{2i-1},
p_{2i})=&2,\qquad p_{2i}=1\cr =&0,\qquad
p_{2i}=0,\ n_{2i}+p_{2i-1}=0\
(\hbox{mod}\ 2)\cr =&4, \qquad
p_{2i}=0,\ n_{2i}+p_{2i-1}=1\
(\hbox{mod}\ 2).}\eqn\serbia$$

The overall moduli and fixed tori
independent normalization is easily obtained from the formula
derived for ${\bf Z}_N$ orbifolds in terms of twists in the
two-dimensional subspaces [\eleven] which also applies here.
\chapter {Conclusions}
The moduli dependence of the Yukawa couplings amongst twisted
sectors has been investigated for all ${\bf Z}_M\times {\bf Z}_N$ orbifolds
with the point group realized in the simplest way by Coxeter
elements acting on a direct sum of two-dimensional root lattices.
All Yukawa couplings have a moduli dependence of the general form
\ale, \ba, \baili\ or \hellraiser.
A tabulation has been made of the moduli dependent suppression
factors for all the non-zero Yukawa couplings. For any definite
model, with specified point group embeddings and Wilson lines, the
results of this paper may be employed to study suppression factors
in the quark and lepton mass matrices.

\centerline{ACKNOWLEDGEMENTS}
This work was supported in part by S.E.R.C. We would like to thank
S. Stieberger for a helpful communication.
\refout
\vfill\eject
\centerline{\bf{Table Captions}}
Table 1: Point group elements and lattices for ${\bf Z}_M\times {\bf Z}_N$
orbifolds. The point group elements $\theta$ and $\omega$ which are
of the form $(e{2\pi i\eta_1},e{2\pi i\eta_2},e{2\pi
i\eta_3})$, are specified in the table by $(\eta_1,
\eta_2,\eta_3).$

Table 2: Fixed points (and tori) for the ${\bf Z}_2\times {\bf Z}_6$ orbifold
and associated space group elements. The inequivalent fixed points
(or tori) are obtained by giving integers $n$ with half integral
coefficients the values $0,1$ and integers $n$ with third integral
coefficients the values $0,\pm 1$. The coefficients $a_1,...,a_6$
are arbitrary, and for the space group elements $(\alpha, l)$ the
lattice vector $(I-\alpha)\Lambda$, where $\Lambda$ is an arbitrary
lattice vector, may be added to $l$.

Table 3: Action of $\theta$ and $\omega$ on the fixed tori for the
${\bf Z}_2\times {\bf Z}_6$ orbifold.

Table 4: Suppression factors $e{-{\lambda_1R_12/ 4\pi \sqrt
3}}$$e{-{\lambda_3R_32/ 4\pi \sqrt
3}}$$e{-{\lambda_5R_52/ 4\pi \sqrt
3}}$ for the ${\bf Z}_2\times {\bf Z}_6$ orbifold Yukawa couplings. The values
of the integers characterizing fixed points (or tori) for which
non-trivial suppression factors occur are indicated.
The $\theta\alpha\omega\beta$ twisted sector is denoted by
$T_{\alpha\beta}$.
\vfill\eject
\centerline{\bf{TABLE 1}}
\vskip 0.5cm
\begintable
Point Group |$\theta$ |$\omega$ |Lattice\cr${\bf Z}_2\times
{\bf Z}_2$ |$(1,0,-1)/2$ |$(0,1,-1)/2$ |$SO(4)3$\cr${\bf Z}_3\times
{\bf Z}_3$ |$(1,0,-1)/3$ |$(0,1,-1)/3$ |$SU(3)3$\cr${\bf Z}_2\times
{\bf Z}_4$ |$(1,0,-1)/2$ |$(0,1,-1)/4$ |$SO(4)\times SO(5)2$\cr${\bf
Z}_4\times
{\bf Z}_4$ |$(1,0,-1)/4$ |$(0,1,-1)/4$ |$SO(5)3$\cr${\bf Z}_2\times
{\bf Z}_6$ |$(1,0,-1)/2$ |$(0,1,-1)/6$ |$SO(4)\times G_22$\cr
${\bf Z}_2\times
{\bf Z}'_6$ |$(1,0,-1)/2$ |$(1,1,-2)/6$ |$G_23$\cr
${\bf Z}_3\times
{\bf Z}_6$ |$(1,0,-1)/3$ |$(0,1,-1)/6$ |$SU(3)\times G_22$ \cr${\bf
Z}_6\times {\bf Z}_6$ |$(1,0,-1)/6$ |$(0,1,-1)/6$ |$G_23$
\endtable
\centerline{\bf{TABLE 2}}
\vskip 0.5cm
\begintable
twisted sector |Fixed point
or torus |$l$\cr$\theta$|${\displaystyle n_1\over\displaystyle
2}e_1+{\displaystyle n_2\over\displaystyle 2} e_2+{\displaystyle
n_5\over\displaystyle 2}e_5+{\displaystyle n_6\over\displaystyle 2}
e_6+a_3 e_3+a_4 e_4$|$n_1e_1+n_2e_2+n_5e_5+n_6e_6$ \cr$\omega$
|$a_1e_1+a_2e_2$ |$0$\cr $\omega2$|$-{\displaystyle
n_3\over\displaystyle  3}e_4+{\displaystyle n_5\over\displaystyle
3}e_6+a_1e_1+a_2e_2$|$n_3e_3+n_5e_5$ \cr$\omega3$|$
{\displaystyle n_3\over\displaystyle  2}e_3+{\displaystyle
n_4\over\displaystyle 2}e_4+{\displaystyle n_5\over\displaystyle
2}e_5+{\displaystyle n_6\over\displaystyle
2}e_6+a_1e_1+a_2e_2$|$n_3e_3+n_4e_4+n_5e_5+n_6e_6$
\cr$\omega4$|$-{\displaystyle n_3\over\displaystyle
3}e_4+{\displaystyle n_5\over\displaystyle
3}e_6+a_1e_1+a_2e_2$|$-n_3e_3-n_5e_5$\cr
$\omega5$|$a_1e_1+a_2e_2$|$0$ \cr$\theta\omega$ |$-{\displaystyle
n_1\over\displaystyle 2}e_1-{\displaystyle n_2\over\displaystyle
2}e_2-{\displaystyle n_5\over\displaystyle 3}e_6$ |
$n_1e_1+n_2e_2+n_5e_5$\cr$\theta\omega2$ |$-{\displaystyle
n_1\over\displaystyle 2}e_1-{\displaystyle n_2\over\displaystyle
2}e_2-{\displaystyle n_3\over\displaystyle 3}e_4$|
$n_1e_1+n_2e_2+n_3e_3$ \cr $\theta\omega3$| ${\displaystyle
n_1\over\displaystyle 2}e_1+{\displaystyle n_2\over\displaystyle
2}e_2+{\displaystyle n_3\over\displaystyle 2}e_3+ {\displaystyle
n_4\over\displaystyle
2}e_4+a_5e_5+a_6e_6$|$n_1e_1+n_2e_2+n_3e_3+n_4e_4$ \endtable
\vfill\eject \centerline{\bf{TABLE 3}} \vskip 0.5cm \begintable
twisted sector |Action of $\theta$
 |Action of $\omega$\cr$\theta$
 |$I$ |$(n_5,n_6)\rightarrow (n_6,n_5+n_6)$
\cr$\omega$ |$I$ |$I$
\cr$\omega2$ |$n_5\rightarrow -n_5$
 |$(n_3,n_5)\rightarrow (-n_3,-n_5)$
\cr$\omega3$ |$I$ |$(n_3,n_4,n_5,n_6)\rightarrow
(n_3+n_4,n_3,n_6,n_5+n_6)$\cr
$\omega4$
 |$n_5\rightarrow -n_5$ |
$(n_3,n_5)\rightarrow (-n_3,-n_5)$
\cr$\omega5$ |$I$ |$I$
\cr$\theta\omega$ |$n_5\rightarrow -n_5$ |
$I$\cr$
\theta\omega2$ |$I$ |
$n_3\rightarrow -n_3$
\cr$\theta\omega3$ |
$I$ |
$(n_3,n_4)\rightarrow
(n_4,n_3+n_4)$
\endtable
\centerline{\bf{TABLE 4}}
\vskip 0.5cm
\begintable
Yukawa
Coupling|$\lambda_1$|$\lambda_2$|$\lambda_3$\cr
$T_{01}T_{12}T_{13}$|0|$\mu(n_3,p_3,p_4)$|0\cr
$T_{02}T_{11}T_{13}$|0|$\mu(n_3,p_3,p_4)$|0\cr
$T_{03}T_{10}T_{13}$|0|0|0\cr
$T_{04}T_{10}T_{12}$|0|0|$\mu(n_5,p_5,p_6)$\cr
$T_{04}T_{11}T_{11}$|0|1 for $n_3(T_{04})\not=0$|2 for
$m_5(T_{11})-n_5(T_{11})\not=0$\cr
$T_{05}T_{10}T_{11}$|0|0|$\mu(n_5,p_5,p_6)$
\endtable
\vfill\eject
\centerline{\bf{APPENDIX}}\hfill\break
$\underline {{\bf Z}_2\times {\bf Z}_2\ Orbifold}$\hfill\break
$\underline{Lattice:\ SO(4)3}$\hfill\break $$\theta=(C(SO(4)), I,
C(SO(4))),\quad \omega=(I, C(SO(4)), C(SO(4))).$$
The fixed tori and associated space group elements are given with
the conventions of table 2 by
\vskip 0.5cm
\begintable
twisted sector| Fixed point
or torus| $l$\cr
$\theta$ |${\displaystyle n_1\over\displaystyle
2}e_1+{\displaystyle n_2\over\displaystyle 2}
e_2+{\displaystyle n_5\over\displaystyle 2}e_5+{\displaystyle
n_6\over\displaystyle 2}e_6+a_3e_3+a_4e_4$|
$n_1e_1+n_2e_2+n_5e_5+n_6e_6$\cr $\omega$|${\displaystyle
n_3\over\displaystyle 2}e_3+{\displaystyle n_4\over\displaystyle 2}
e_4+{\displaystyle n_5\over\displaystyle
2}e_5+{\displaystyle n_6\over\displaystyle 2}e_6+a_1
e_1+a_2e_2$|$n_3e_3+n_4e_4+n_5e_5+n_6e_6$\cr
$\theta\omega$|${\displaystyle n_1\over\displaystyle
2}e_1+{\displaystyle n_2\over\displaystyle 2} e_2+{\displaystyle
n_3\over\displaystyle 2}e_3+{\displaystyle n_4\over\displaystyle
2}e_4+a_5e_5+a_6e_6$| $n_1e_1+n_2e_2+n_3e_3+n_4e_4$ \endtable  The
only non-zero Yukawa couplings is $T_{01}T_{10}T_{11}$, where
$T_{\alpha\beta}$ denotes the $\theta\alpha\omega\beta$ twisted
sector, and since no complex plane is twisted by all three twist
fields  there is no moduli dependence.\hfill\break  $\underline{{\bf
Z}_3\times {\bf Z}_3\ Orbifold}$\hfill\break  $\underline{Lattice\
SU(3)3}$\hfill\break
Already discussed in [\twelve].\hfill\break
\hfill\break
$\underline{{\bf Z}_2\times {\bf Z}_4\ Orbifold}$\hfill\break
$\underline{lattice\ SO(4)\times SO(5)\times SO(5)}$\hfill\break
In the orthonormal basis,
$$\eqalign{&\theta X_1=-X_1,\cr &\theta
X_2=X_2,\cr &\theta X_3=-X_3,}\qquad
\eqalign{&\omega X_1=X_1,\cr &\omega X_2=e{2\pi i/4}X_2,\cr &\omega
X_3=e{-2\pi i/4}X_3.}$$
In the lattice basis,$$\theta=(C(SO(4)), I, C{2}(SO(5))),\quad
\omega=(I, C(SO(5)),C{-1}(SO(5)))$$ so that
$$\eqalign{&\theta e_1=-e_1,\cr
&\theta e_2=-e_2,}\qquad\eqalign{&\theta e_3=e_3,\cr
&\theta e_4=e_4,}\qquad\eqalign{&\theta e_5=-e_5,\cr &\theta
e_6=-e_6}$$
and $$\eqalign{&\omega e_1=e_1,\cr &\omega
e_2=e_2,}\qquad\eqalign{&\omega e_3=e_3+2e_4,\cr &\omega
e_4=-e_3-e_4,}\qquad\eqalign{&\omega e_5=-e_5-2e_6,\cr &\omega
e_6=e_5+e_6.}$$
The independent deformation parameters (moduli) are
$R_1$, $R_2$, $\cos\theta_{12}$, $R_3$ and $R_5$, where
$$\vert e_1\vert\equiv R_1,\quad \vert e_2\vert\equiv R_2,\quad
\vert e_3\vert=\sqrt 2\vert e_4\vert\equiv R_3,\quad \vert
e_5\vert=\sqrt 2\vert e_6\vert\equiv R_5,$$ also,
$$\cos{\theta_{34}}=\cos{\theta_{56}}=-{1\over\displaystyle  \sqrt2}$$ with all
other $\cos{\theta_{ij}}$ except $\cos{\theta_{12}}$ zero. A
suitable choice of orthonormal basis $\tilde e_1,...,\tilde e_6$ is
given by
$$\eqalign{&e_1=R_1\tilde e_1,\cr
&e_2=R_2(\cos\theta_{12}\tilde e_1+\sin\theta_{12}\tilde
e_2),}\quad\eqalign{&e_3=R_3\tilde e_3,\cr
&e_4={R_3\over2}(-\tilde e_3+\tilde
e_4),}\quad\eqalign{&e_5=R_5\tilde e_5,\cr
&e_6={R_5\over2}(-\tilde
e_5+\tilde e_6).}$$   The fixed points or
tori and associated space group elements are given with the
conventions of table 2 by \vskip 0.5cm \begintable twisted
sector|Fixed point or torus|$l$\cr $\theta$ |${\displaystyle
n_1\over\displaystyle 2}e_1+{\displaystyle n_2\over\displaystyle
2}e_2+ {\displaystyle n_5\over\displaystyle 2}e_5+{\displaystyle
n_6\over\displaystyle 2}e_6+a_3e_3+a_4e_4$
|$n_1e_1+n_2e_2+n_5e_5+n_6e_6$ \cr
$\omega$|$a_1e_1+a_2e_2-{\displaystyle n_4\over\displaystyle 2}e_3+
{\displaystyle n_6\over\displaystyle 2}e_5$|$n_4e_4-n_6e_6$ \cr
$\omega2$|$a_1e_1+a_2e_2+
{\displaystyle n_3\over\displaystyle
2}e_3+{\displaystyle n_4\over\displaystyle 2}e_4+ {\displaystyle
n_5\over\displaystyle 2}e_5+{\displaystyle n_6\over\displaystyle
2}e_6$| $n_3e_3+n_4e_4+n_5e_5+n_6e_6$\cr
$\omega3$|$a_1e_1+a_2e_2-{\displaystyle n_4\over\displaystyle
2}e_3+{\displaystyle n_6\over\displaystyle
2}e_5$|$-n_4e_4+n_6e_6$\cr $\theta\omega$|$ {\displaystyle
n_1\over\displaystyle 2}e_1+{\displaystyle n_2\over\displaystyle
2}e_2-{\displaystyle n_4\over\displaystyle
2}e_3+{\displaystyle n_6\over\displaystyle 2}e_5$|
$n_1e_1+n_2e_2+n_4e_4+n_6e_6$\cr $\theta\omega2$|$ {\displaystyle
n_1\over\displaystyle 2}e_1+ {\displaystyle n_2\over\displaystyle
2}e_2+{\displaystyle n_3\over\displaystyle 2}e_3+{\displaystyle
n_4\over\displaystyle 2}e_4+ a_5e_5+a_6e_6$|
$n_1e_1+n_2e_2+n_3e_3+n_4e_4$\cr $\theta\omega3$|$ {\displaystyle
n_1\over\displaystyle 2}e_1+{\displaystyle n_2\over\displaystyle
2}e_2-{\displaystyle n_4\over\displaystyle 2}e_3+{\displaystyle
 n_6\over\displaystyle 2}e_5$|
$n_1e_1+n_2e_2+n_6e_6-n_4e_4$ \endtable
The action of
$\theta$ and $\omega$ on the fixed points (or tori) is given by
\vskip 0.5cm
\begintable
Twisted sector|Action of $\theta$|Action of $\omega$\cr
$\theta$|$I$|$n_5\rightarrow n_5+n_6$\cr
$\omega,\omega3$|$I$|$I$\cr
$\omega2$|$I$|$(n_3,n_5)\rightarrow (n_3+n_4,n_5+n_6)$ \cr
$\theta\omega,\theta\omega3$|$I$|$I$ \cr
$\theta\omega2$|$I$|$n_3\rightarrow n_3+n_4$
\endtable
The moduli dependent suppression factors for the non-zero Yukawa
couplings amongst twisted sectors containing massless states
written in the form $e{-\lambda_{2i-1}R_{2i-1}2/16\pi}$ are given
by
\vskip 0.5cm
\begintable
Yukawa
Coupling|$\lambda_1$|$\lambda_2$|$\lambda_3$\cr
$T_{01}T_{11}T_{12}$|0|$\rho(n_4,p_3,p_4)$|0\cr
$T_{02}T_{10}T_{12}$|0|0|0\cr
$T_{03}T_{10}T_{11}$|0|0|$\rho(n_6,p_5,p_6)$\cr
$T_{11}T_{11}T_{02}$|0|$\rho(n_4,p_3,p_4)$|
$\rho(n_6,p_5,p_6)$
\endtable
\vfill\eject\hfill\break
$\underline{{\bf Z}_4\times {\bf
Z}_4\ Orbifold}$\hfill\break  $\underline{Lattice\
SO(5)\times SO(5)\times SO(5)}$\hfill\break
In the orthonormal basis,
$$\eqalign{&\theta X_1=e{2\pi i/4}X_1,\cr &\theta
X_2=X_2,\cr &\theta X_3=e{-2\pi i/4}X_3,}\qquad
\eqalign{&\omega X_1=X_1,\cr &\omega X_2=e{2\pi i/4}X_2,\cr &\omega
X_3=e{-2\pi i/4}X_3.}$$
In the lattice basis,
$$\theta=(C(SO(5)), I, C{-1}(SO(5))),\quad \omega=(I,
C(SO(5)),C{-1}(SO(5)))$$ so that
$$\eqalign{&\theta e_1=e_1+2e_2,\cr &\theta e_2=-e_1-e_2,}\qquad
\eqalign{&\theta e_3=e_3,\cr &\theta e_4=e_4,}\qquad
\eqalign{&\theta e_5=-e_5-2e_6,\cr &\theta e_6=e_5+e_6}$$
and
$$\eqalign{&\omega e_1=e_1,\cr &\omega
e_2=e_2,}\qquad\eqalign{&\omega e_3=e_3+2e_4,\cr &\omega
e_4=-e_3-e_4,}\qquad \eqalign{&\omega e_5=-e_5-2e_6,\cr &\omega
e_6=e_5+e_6.}$$
The independent deformation parameters (moduli) are
$R_1$, $R_3$ and $R_5$, where
$$ \vert e_1\vert=\sqrt2 \vert e_2\vert\equiv R_1,
\quad \vert e_3\vert =\sqrt2\vert e_4\vert\equiv R_3,\quad
\vert e_5\vert=\sqrt 2\vert e_6\vert\equiv R_5,$$
also$$\cos\theta_{12}=\cos\theta_{34}=\cos\theta_{56}=-{1\over\sqrt2}$$
with all other $\cos\theta_{ij}$ zero. A suitable choice of
orthonormal basis $\tilde e_1,..., \tilde e_6$ is given by
$$\eqalign{&e_1=R_1\tilde e_1,\cr
&e_2=-{R_1\over2}(\tilde e_1-\tilde e_2),}\quad\eqalign{&e_3=
R_3\tilde e_3,\cr &e_4=-{R_3\over2}(\tilde
e_3-\tilde e_4),}\quad\eqalign{&e_5=R_5\tilde
e_5,\cr &e_6=-{R_5\over2}(\tilde e_5-\tilde e_6).}$$
\vfill\break
The fixed points or tori and associated space group
elements are given with the conventions of table 2 by
\vskip 0.5cm
\begintable
twisted sector|Fixed point
or torus|$l$\cr
$\theta$|$-{\displaystyle
n_2\over\displaystyle2}e_1+{\displaystyle n_6\over\displaystyle
2}e_5+a_3e_3+a_4 e_4$|$n_2e_2+n_6e_6$\cr
$\omega$|$a_1e_1+a_2e_2-{\displaystyle n_4\over\displaystyle
2}e_3+{\displaystyle n_6\over\displaystyle 2}e_5$|$n_4e_4+n_6e_6$\cr
$\omega2$|$a_1e_1+a_2e_2+{\displaystyle
n_3\over\displaystyle2}e_3+{\displaystyle
n_4\over\displaystyle2}e_4+{\displaystyle
n_5\over\displaystyle 2}e_5+{\displaystyle n_6\over\displaystyle
2}e_6$|$n_3e_3+n_4e_4+n_5e_5+n_6e_6$\cr
$\omega3$|$a_1e_1+a_2e_2+{\displaystyle
n_4\over\displaystyle 2}e_3-{\displaystyle n_6\over\displaystyle
2}e_5$|$n_4e_4+n_6e_6$\cr
$\theta\omega$|$-{\displaystyle
n_2\over\displaystyle2}e_1-{\displaystyle n_4\over\displaystyle2}
e_3+{\displaystyle n_5\over\displaystyle2}e_5+{\displaystyle
n_6\over\displaystyle 2}e_6$|$n_2e_2+n_4e_4+n_5e_5+n_6e_6$\cr
$\theta\omega2$|$-{\displaystyle n_2\over\displaystyle 2}
e_1+{\displaystyle n_3\over\displaystyle 2} e_3+{\displaystyle
n_4\over\displaystyle 2}e_4-{\displaystyle n_6\over\displaystyle
2}e_5$|  $n_2e_2+n_4e_4+n_3e_3+n_6e_6$\cr
$\theta\omega3$|$-{\displaystyle n_2\over2} e_1+{\displaystyle
n_4\over\displaystyle 2} e_3+a_5 e_5+a_6e_6$| $n_2e_2+n_4e_4$\cr
$\theta2$|${\displaystyle n_1\over\displaystyle
2}e_1+{\displaystyle n_2\over\displaystyle 2} e_2+a_3 e_3+a_4
e_4+{\displaystyle n_5\over\displaystyle 2}e_5+{\displaystyle
n_6\over\displaystyle 2}e_6$|  $n_1e_1+n_2e_2+n_5e_5+n_6e_6$\cr
$\theta3$| ${\displaystyle n_2\over2}e_1+a_3e_3+a_4e_4-
{\displaystyle n_6\over\displaystyle2}e_5$|
$n_2e_2+n_6e_6$\cr $\theta2\omega$|${\displaystyle
n_1\over\displaystyle 2}e_1+{\displaystyle n_2\over\displaystyle 2}
e_2- {\displaystyle n_4\over\displaystyle 2}e_3-{\displaystyle
n_6\over\displaystyle 2}e_5$|  $n_1e_1+n_2e_2+n_4e_4+n_6e_6$\cr
$\theta2\omega2$|   ${\displaystyle n_1\over\displaystyle2}e_1+{\displaystyle
n_2\over\displaystyle2} e_2+{\displaystyle
n_3\over\displaystyle 2}e_3+{\displaystyle n_4\over\displaystyle
2}e_4+a_5 e_5+a_6e_6$|  $n_1e_1+n_2e_2+n_3e_3+n_4e_4$\cr
$\theta3\omega$|${\displaystyle n_2\over\displaystyle
2}e_1-{\displaystyle n_4\over\displaystyle 2} e_3+a_5e_5+a_6 e_6$|
$n_2e_2+n_4e_4$\cr
$\theta2\omega3$|
${\displaystyle
n_1\over\displaystyle 2}e_1+{\displaystyle n_2\over\displaystyle 2}
e_2+{\displaystyle n_4\over\displaystyle 2}e_3+{\displaystyle
n_6\over\displaystyle 2}e_5$|
$n_1e_1+n_2e_2+n_4e_4+n_6e_6$\cr
$\theta3\omega2$|${\displaystyle
n_2\over\displaystyle 2}e_1+{\displaystyle n_3\over\displaystyle 2}
e_3+{\displaystyle n_4\over\displaystyle 2}e_4+{\displaystyle
n_6\over\displaystyle 2}e_5$|  $n_2e_2+n_3e_3+n_4e_4+n_6e_6$
\cr
$\theta3\omega3$|${\displaystyle
n_2\over\displaystyle 2}e_1+{\displaystyle n_4\over\displaystyle 2}
e_3+{\displaystyle n_5\over\displaystyle 2}e_5+{\displaystyle
n_6\over\displaystyle 2}e_6$|  $n_2e_2+n_4e_4+n_5e_5+n_6e_6$
\endtable
\vfill\eject
The action of
$\theta$ and $\omega$ on the fixed points (or tori) is given by
\vskip 0.5cm
\begintable
twisted sector| Action of theta|Action of omega\cr
$\theta,\theta3$|$I$|$I$\cr
$\omega,\omega3$|$I$|$I$\cr
$\theta3\omega,\theta\omega3$|$I$|$I$\cr
$\theta\omega,\theta3\omega3$|$n_5\rightarrow
n_5+n_6$|$n_5\rightarrow n_5+n_6$\cr
$\theta\omega2,\theta3\omega2$|$I$|$n_3\rightarrow n_3+n_4$\cr
$\theta2\omega, \theta2\omega3$|$n_1\rightarrow
n_1+n_2$|$I$\cr
$\omega2$|$n_5\rightarrow n_5+n_6$|$(n_3,n_5)\rightarrow
(n_3+n_4,n_5+n_6)$\cr
$\theta2$|$(n_1,n_5)\rightarrow (n_1+n_2,
n_5+n_6)$|$n_5\rightarrow n_5+n_6$\cr
$\theta2\omega2$|$n_1\rightarrow
n_1+n_2$|$n_3\rightarrow n_3+n_4$
\endtable
\vfill\eject
The moduli dependent suppression factors for the non-zero
Yukawa couplings amongst twisted sectors containing massless states
written in the form $e{-\lambda_{2i-1}R_{2i-1}2/16\pi}$ are given
by
\vskip 0.5cm
\begintable
Yukawa coupling|$\lambda_1$|$\lambda_2$|
$\lambda_3$\cr
$T_{01}T_{12}T_{31}$|0|$\rho(n_4,p_3,p_4)$|0\cr
$T_{01}T_{13}T_{30}$|0|0|0\cr
$T_{01}T_{21}T_{22}$|0|$\rho(n_4,p_3,p_4)$|0\cr
$T_{02}T_{11}T_{31}$|0|$\rho(n_4,p_3,p_4)$|0\cr
$T_{02}T_{12}T_{30}$|0|0|$\rho(n_6,p_5,p_6)$\cr
$T_{02}T_{20}T_{22}$|0|0|0\cr
$T_{02}T_{21}T_{21}$|0|$\rho(n_4,p_3,p_4)$|
$\rho(n_6,p_5,p_6)$\cr
$T_{03}T_{10}T_{31}$|0|0|0\cr
$T_{03}T_{11}T_{30}$|0|0|$\rho(n_6,p_5,p_6)$\cr
$T_{03}T_{20}T_{21}$|0|0|$\rho(n_6,p_5,p_6)$\cr
$T_{10}T_{12}T_{22}$|$\rho(n_2,p_1,p_2)$|0|0\cr
$T_{10}T_{13}T_{21}$|$\rho(n_2,p_1,p_2)$|0|0\cr
$T_{11}T_{12}T_{21}$|$\rho(n_2,p_1,p_2)$|$\rho(n_4,p_3,p_4)$|
$\rho(n_6,p_5,p_6)$\cr
$T_{11}T_{11}T_{22}$|$\rho(n_2,p_1,p_2)$|$\rho(n_4,p_3,p_4)$|0\cr

$T_{11}T_{13}T_{20}$|$\rho(n_2,p_1,p_2)$|0|0\cr
$T_{12}T_{12}T_{20}$|$\rho(n_2,p_1,p_2)$|0|$\rho(n_6,p_5,p_6)$
\endtable
\hfill\eject\hfill\break
$\underline{{\bf Z}_2\times {\bf
Z}_6\ Orbifold}$\hfill\break  $\underline{Lattice\
SO(4)\times G_22}$\hfill\break
Discussed in section 3.\hfill\break
$\underline{{\bf Z}_2\times {\bf
Z}'_6\ Orbifold}$\hfill\break  $\underline{Lattice\
G_23}$\hfill\break In the orthonormal basis,
$$\eqalign{&\theta X_1=-X_1,\cr &\theta X_2=X_2,\cr &\theta
X_3=-X_3,}\qquad\eqalign{&\omega X_1=e{2\pi i/6}X_1,\cr &\omega
X_2=e{2\pi i/6}X_2,\cr &\omega X_3=e{-2\pi i/3}X_3.}$$
In the lattice basis,
$$\theta=(C3(G_2), I, C3(G_2)),\quad \omega=(C(G_2), C(G_2),
C{-2}(G_2))$$
so that
$$\eqalign{&\theta e_1=-e_1,\cr & \theta
e_2=-e_2,}\qquad\eqalign{&\theta e_3=e_3,\cr & \theta
e_4=e_4,}\qquad\eqalign{&\theta e_5=-e_5,\cr &\theta e_6=-e_6.}$$
and $$\eqalign{&\omega e_1=-e_1-e_2,\cr
&\omega e_2=3e_1+2e_2,}\qquad\eqalign{&\omega e_3=-e_3-e_4,\cr
&\omega e_4=3e_3+2e_4,}\qquad\eqalign{&\omega e_5=e_5+ e_6,\cr
&\omega e_6=-3e_5-2e_6.}$$
The independent deformation parameters (moduli) are
$R_1$, $R_3$ and $R_5$, where
$$\vert e_2\vert=\sqrt 3\vert e_1\vert\equiv\sqrt 3 R_1,\quad \vert
e_4\vert=\sqrt 3\vert e_3\vert\equiv\sqrt 3 R_3,\quad
\vert e_6\vert=\sqrt 3\vert e_5\vert\equiv\sqrt 3 R_5$$
also,
$$\cos{\theta_{12}}=\cos{\theta_{34}}=\cos{\theta_{56}}=-{\sqrt
3\over\displaystyle  2}$$ with all other $\cos{\theta_{ij}}$ zero.
A suitable choice of
orthonormal basis $\tilde e_1,...,\tilde e_6$ is given by
$$\eqalign{&e_1=R_1\tilde e_1\cr &e_2=R_1(-{3\over\displaystyle  2}\tilde
e_1-{\sqrt 3\over\displaystyle 2}\tilde e_2)}\quad
\quad\eqalign{&e_3=R_3\tilde e_3 \cr &e_4=R_3(-{3\over\displaystyle
2}\tilde e_3-{\sqrt 3\over\displaystyle 2}\tilde
e_4)}\quad \eqalign{&e_5=R_5\tilde e_5\cr &e_6=R_5(-{3\over\displaystyle
2}\tilde e_5-{\sqrt 3\over\displaystyle 2}\tilde
e_6)}$$
The fixed points or tori and associated space group elements are
given with the conventions of table 2 by
\vskip 0.5cm
\begintable
Twisted sector |Fixed point
or torus |$l$\cr
$\theta$|${\displaystyle n_1\over\displaystyle 2}e_1+{\displaystyle
 n_2\over\displaystyle 2}
e_2+{\displaystyle n_5\over\displaystyle 2}e_5+{\displaystyle
 n_6\over\displaystyle 2} e_6+a_3 e_3+a_4
e_4$|$n_1e_1+n_2e_2+n_5e_5+n_6e_6$
\cr$\omega$|${\displaystyle n_5\over\displaystyle 3}e_6$|$n_5e_5$
\cr$\omega2$|$-{\displaystyle n_1\over\displaystyle
3}e_2-{\displaystyle n_3\over\displaystyle 3}e_4+{\displaystyle
 n_5\over\displaystyle 3}e_6$
|$n_1e_1+n_3e_3-n_5e_5$\cr
$\omega3$|${\displaystyle n_1\over\displaystyle
2}e_1+{\displaystyle n_2\over\displaystyle 2}e_2+{\displaystyle
 n_3\over\displaystyle
2}e_3+{\displaystyle n_4\over\displaystyle 2}e_4+a_5e_5+a_6e_6$|
$n_1e_1+n_2e_2+n_3e_3+n_4e_4$
\cr
$\omega4$|$-{\displaystyle n_1\over\displaystyle  3}e_2-{\displaystyle
 n_3\over\displaystyle  3}e_4+{\displaystyle n_5\over\displaystyle 3}e_6
$|$-n_1e_1-n_3e_3+n_5e_5$\cr$
\omega5$|${\displaystyle n_5\over\displaystyle 3}e_6$|$-n_5e_5$\cr
$\theta\omega$|${\displaystyle n_1\over\displaystyle 3}e_1$ |$n_1e_1$\cr
$\theta\omega2$
|$-{\displaystyle n_3\over\displaystyle 3}e_4$|$n_3e_3$\cr$\theta\omega3$|
${\displaystyle n_3\over\displaystyle 2}e_3+{\displaystyle
n_4\over\displaystyle
 2}e_4+{\displaystyle n_5\over\displaystyle 2}e_5+{\displaystyle
 n_6\over\displaystyle 2}e_6+
a_1e_1+a_2e_2$
|$n_3e_3+n_4e_4+n_5e_5+n_6e_6$\cr$
\theta\omega4$|$-{\displaystyle n_3\over\displaystyle 3}e_4$|$-n_3e_3
$\cr$\theta\omega5$|${\displaystyle n_1\over\displaystyle 3}e_2$|$-n_1e_1$
\endtable
\vfill\eject
The action of $\theta$ and $\omega$ on the fixed points
(or tori) is given by
\vskip 0.5cm
\begintable
Twisted sector |Action of $\theta$|Action of $\omega$\cr
$\theta$|$I$|$(n_1,n_2,n_5,n_6)\rightarrow (n_1+n_2,n_1,n_5+n_6,n_5)$
\cr$\omega,\omega5$|$n_5\rightarrow -n_5$|$I$
\cr$\omega2,\omega4$|$(n_3,n_5)\rightarrow
(-n_3,-n_5)$|$(n_1,n_3)\rightarrow (-n_1,-n_3)$\cr
$\omega3$|$I$|$(n_1,n_2,n_3,n_4)\rightarrow (n_1+n_2,n_1,n_3+n_4,n_3)$
\cr$\theta\omega,\theta\omega5$|$n_1\rightarrow -n_1$|
$n_1\rightarrow -n_1$
\cr$
\theta\omega2,\theta\omega4$|$I$|
$n_3\rightarrow -n_3$\cr
$\theta\omega3$|$I$|
$(n_3,n_4,n_5,n_6)\rightarrow
(n_3+n_4,n_3,n_5+n_6,n_5)$
\endtable
\vskip 0.5cm
The moduli dependent suppression factors for the non-zero
Yukawa couplings amongst twisted sectors containing massless states
are given in the conventions of table 4 by
\vskip 0.5cm
\begintable
Yukawa|$\lambda_1$|$\lambda_2$|$\lambda_3$\nr
Coupling| |   |\cr
$T_{01}T_{02}T_{03}$|$\mu(n_1,p_1,p_2)$|$\mu(n_3,p_3,p_4)$|0\cr
$T_{01}T_{11}T_{14}$|1 for $n_1(T_{11})\not=0$|1 for
$n_3(T_{14})\not=0$|1 for
$n_5(T_{01})\not=0$\cr
$T_{02}T_{02}T_{02}$|2 for
$m_1(T_{02})-n_1(T_{02})\not=0$| 2 for
$m_3(T_{02})-n_3(T_{02})\not=0 $|2 for
$m_5(T_{02})-n_5(T_{02})\not=0$\cr
$T_{02}T_{10}T_{14}$|
$\mu(n_1,p_1,p_2)$|0|$\mu(n_5,p_5,p_6)$\cr
$T_{02}T_{11}T_{13}$|0|$\mu(n_3,p_3,p_4)$|$\mu(n_5,p_5,p_6)$\cr
$T_{03}T_{10}T_{13}$|0|0|0
\endtable
\vfill\eject\hfill\break
$\underline{{\bf Z}_3\times {\bf Z}_6\ Orbifold}$\hfill\break
$\underline{lattice\ SU(3)\times G_22}$\hfill\break
In the orthonormal basis,
$$\eqalign{&\theta X_1=e{2\pi i/3}X_1,\cr &\theta
X_2=X_2,\cr &\theta X_3=e{-2\pi i/3}X_3,}\qquad
\eqalign{&\omega X_1=X_1,\cr &\omega X_2=e{2\pi i/6}X_2,\cr &\omega
X_3=e{-2\pi i/6}X_3.}$$
In the lattice basis,
$$\theta=(C(SU(3)), I, C{-2}(G_2)),\quad \omega=(I, C(G_2),
C{-1}(G_2))$$ so that
$$\eqalign{&\theta e_1=e_2,\cr &\theta e_2=-e_1-e_2,}\qquad
\eqalign{&\theta e_3=e_3,\cr &\theta e_4=e_4,}\qquad\eqalign{&\theta
e_5=e_5+e_6,\cr &\theta e_6=-3e_5-2e_6.}$$
and
$$\eqalign{&\omega e_1=e_1,\cr &\omega e_2=e_2,}\qquad\eqalign{&
\omega e_3=-e_3-e_4,\cr &\omega e_4=3e_3+2e_4,}\qquad
\eqalign{&\omega e_5=2e_5+
e_6,\cr & \omega e_6=-3e_5-e_6.}$$
 The independent deformation parameters (moduli) are
$R_1$, $R_3$ and $R_5$, where
$$\vert e_2\vert=\vert e_1\vert\equiv R_1,\quad \vert
e_4\vert=\sqrt 3\vert e_3\vert\equiv\sqrt 3 R_3,\quad
\vert e_6\vert=\sqrt 3\vert e_5\vert\equiv\sqrt 3 R_5$$
also,
$$\cos{\theta_{12}}=-{1\over\displaystyle 2},\quad
\cos{\theta_{34}}=\cos{\theta_{56}}=-{\sqrt 3\over\displaystyle  2}$$
with all other $\cos{\theta_{ij}}$ zero.
A suitable
choice of orthonormal basis $\tilde e_1,...,\tilde e_6$ is given by
$$\eqalign{&e_1=R_1\tilde e_1,\cr & e_2=R_1(-{1\over\displaystyle
2}\tilde e_1+{\sqrt 3\over\displaystyle 2}\tilde e_2),}\quad
\eqalign{&e_3=R_3\tilde e_3,\cr &e_4=R_3(-{3\over\displaystyle
2}\tilde e_3-{\sqrt 3\over\displaystyle 2}\tilde e_4),}\quad
\eqalign{&e_5=R_5\tilde e_5, \cr &e_6=R_5(-{3\over\displaystyle
2}\tilde e_5-{\sqrt 3\over\displaystyle 2}\tilde e_6).}$$
The fixed points or tori and associated group elements are given
with the conventions of table 2 by
\vfill\eject
\begintable
Twisted sector|Fixed point
or torus|$l$\cr
$\theta$ |${\displaystyle n_1\over\displaystyle
3}(2e_1+e_2)+{\displaystyle n_5\over\displaystyle 3}e_6+a_3e_3+a_4
e_4$|$n_1e_1+n_5e_5$\cr  $\theta2$|${\displaystyle
n_1\over\displaystyle3}(2e_1+e_2)+{\displaystyle
n_5\over\displaystyle  3}e_6+a_3e_3+ a_4 e_4$ |$-n_1e_1-n_5e_5$\cr
$\omega$|$a_1e_1+a_2e_2$|0\cr $\omega2$|$-{\displaystyle
n_3\over\displaystyle 3}e_4+{\displaystyle n_5\over\displaystyle
3}e_6+a_1e_1+a_2e_2$|$n_3e_3+n_5e_5$\cr $\omega3$|${\displaystyle
n_3\over\displaystyle  2}e_3+{\displaystyle n_4\over\displaystyle
2}e_4+{\displaystyle n_5\over\displaystyle  2}e_5+{\displaystyle
n_6\over\displaystyle 2}e_6+a_1e_1+a_2 e_2$|$n_3e_3+n_4e_4+n_5
e_5+n_6e_6$\cr $\omega4$|$-{\displaystyle n_3\over\displaystyle
3}e_4+{\displaystyle n_5\over\displaystyle  3}e_6+a_1e_1+a_2
e_2$|$-n_3e_3-n_5e_5$\cr
$\omega5$|$a_1e_1+a_2e_2$|0
\cr
$\theta\omega$|${\displaystyle n_1\over\displaystyle
3}(2e_1+e_2)+{\displaystyle n_5\over\displaystyle
2}e_5+{\displaystyle n_6\over\displaystyle 2}e_6 $|$n_1e_1+n_5e_5+
n_6e_6$\cr$ \theta\omega2$|${\displaystyle n_1\over\displaystyle
3}(2e_1+e_2)-{\displaystyle n_3\over\displaystyle
3}e_4+{\displaystyle n_5\over\displaystyle 3}e_6 $|$n_1e_1+n_3e_3-
n_5e_5$\cr$ \theta\omega3$|${\displaystyle n_1\over\displaystyle
3}(2e_1+e_2)+{\displaystyle n_3\over\displaystyle
2}e_3+{\displaystyle n_4\over\displaystyle 2}e_4 $|$n_1e_1+n_3e_3+
n_4e_4$\cr $\theta\omega4$|$
{\displaystyle n_1\over\displaystyle 3}(2e_1+e_2)-{\displaystyle
n_3\over\displaystyle 2}e_4+a_5e_5+a_6e_6$|$n_1e_1-n_3e_3 $\cr
$\theta\omega5$|${\displaystyle n_1\over\displaystyle 3}(2e_1+e_2)$
|$n_1e_1$\cr$\theta2\omega$|${\displaystyle n_1\over\displaystyle
3}(2e_1+e_2)$|$-n_1e_1$\cr$ \theta2\omega2$|${\displaystyle
n_1\over\displaystyle 3}(2e_1+e_2)-{\displaystyle
n_3\over\displaystyle 3}e_4 $|$-n_1e_1+n_3e_3$\cr
$\theta2\omega3$|$ {\displaystyle n_1\over\displaystyle
3}(2e_1+e_2)+{\displaystyle n_3\over\displaystyle
2}e_3+{\displaystyle n_4\over\displaystyle 2}e_4$|
$-n_1e_1+n_3e_3+n_4e_4$\cr $\theta2\omega4$| ${\displaystyle
n_1\over\displaystyle 3}(2e_1+e_2)-{\displaystyle
n_3\over\displaystyle 3}e_4+{\displaystyle n_5\over\displaystyle
3}e_6$| $-n_1e_1-n_3e_3+n_5e_5$\cr $\theta2\omega5$|
${\displaystyle n_1\over\displaystyle 3}(2e_1+e_2)+{\displaystyle
n_5\over\displaystyle 2}e_5+{\displaystyle n_6\over\displaystyle
2}e_6$| $-n_1e_1+n_5e_5+n_6e_6$ \endtable \vfill\eject
The action of $\theta$ and $\omega$ on the fixed points
(or tori) is given by
\vskip 0.5cm
\begintable
Twisted sector |Action of $\theta$|Action of
$\omega$\cr
$\theta,\theta2$|$I$|$n_5\rightarrow -n_5$
\cr$\omega,\omega5$ |$I$|$I$
\cr
$\omega2,\omega4$ |$I$|$(n_3,n_5)\rightarrow (-n_3,-n_5)$
\cr
$\omega3$|$(n_5,n_6)\rightarrow
(n_5+n_6,n_5)$|$(n_3,n_4,n_5,n_6)\rightarrow
(n_3+n_4,n_3,n_6,n_5+n_6)$
\cr$\theta\omega,\theta2\omega5$
 |$(n_5,n_6)\rightarrow
(n_5+n_6,n_5)$|
$(n_5,n_6)\rightarrow
(n_6,n_5+n_6)$\cr
$\theta\omega2,\theta2\omega4$|$I$|$(n_3,n_5)\rightarrow
(-n_3,-n_5)$\cr
$\theta\omega3,\theta2\omega3$|$I$|$(n_3,n_4)\rightarrow
(n_3+n_4,n_3)$\cr
$\theta\omega4,\theta2\omega2$|$I$|$n_3\rightarrow
-n_3$\cr$\theta\omega5,\theta2\omega$|$I$|$I$
\endtable
\vfill\eject
The moduli dependent suppression factors for the non-zero Yukawa
couplings amongst twisted sectors containing massless states are
given in the conventions of table 4 by
\vskip 0.5cm
\begintable
Yukawa|$\lambda_1$|$\lambda_2$|$\lambda_3$\nr
Coupling| | |\cr
$T_{01}T_{13}T_{22}$|0|$\mu(n_3,p_3,p_4)$|0\cr
$T_{01}T_{14}T_{21}$|0|1 for $n_3(T_{14})\not=0$|0\cr
$T_{02}T_{12}T_{22}$|0|
2 for $m_3(T_{02})-n_3(T_{12})\not=0$|0\cr
$T_{02}T_{13}T_{21}$|0|$\mu(n_3,p_3,p_4)$|1 for
$n_5(T_{02})\not=0$\cr $T_{02}T_{14}T_{20}$|0|0|0\cr
$T_{03}T_{11}T_{22}$|0|$\mu(n_3,p_3,p_4)$|0\cr
$T_{03}T_{12}T_{21}$|0|$\mu(n_3,p_3,p_4)$|$\mu(n_5,p_5,p_6)$\cr
$T_{03}T_{13}T_{20}$|0|0|$\mu(n_5,p_5,p_6)$\cr
$T_{04}T_{10}T_{22}$|0|0|0\cr
$T_{04}T_{11}T_{21}$|0|1
for $n_3(T_{04})\not=0$|$\mu(n_5,p_5,p_6)$\cr
$T_{04}T_{12}T_{20}$|0|0|2
for $m_5(T_{04})-n_5(T_{12})\not=0$\cr
$T_{05}T_{10}T_{21}$|0|0|1
for $n_5(T_{10})\not=0$\cr
$T_{05}T_{11}T_{20}$|0|0|$\mu(n_5,p_5,p_6)$\cr
$T_{10}T_{12}T_{14}$|2
for $m_1(T_{10})-n_1(T_{12})\not=0$|0|0\cr
$T_{10}T_{13}T_{13}$|2
for $m_1(T_{10})-n_1(T_{13})\not=0$|0|1
for $n_5(T_{10})\not=0$\cr
$T_{11}T_{11}T_{14}$|2
for $m_1(T_{11})-n_1(T_{11})\not=0$|1
for $n_3(T_{14})\not=0$|0\cr
$T_{11}T_{12}T_{13}$|2
for $m_1(T_{11})-n_1(T_{12})\not=0$|$\mu(n_3,p_3,p_4)$|
$\mu(n_5,p_5,p_6)$\cr
$T_{12}T_{12}T_{12}$|2
for $m_1(T_{12})-n_1(T_{12})\not=0$|2
for $m_3(T_{12})-n_3(T_{12})\not=0$| 2
for $m_5(T_{12})-n_5(T_{12})\not=0$
\endtable
\vfill\eject\hfill\break
$\underline{{\bf Z}_6\times {\bf Z}_6\ Orbifold}$\hfill\break
$\underline {Lattice\ G_23}$\hfill\break
In the orthonormal basis, $$\eqalign{&\theta X_1=e{2\pi
i/6}X_1,\cr &\theta X_2=X_2,\cr &\theta X_3=e{-2\pi i/6}X_3,}\quad
\eqalign{&\omega X_1=X_1,\cr &\omega X_2=e{2\pi i/6}X_2,\cr &\omega
X_3=e{-2\pi i/6}X_3.}$$    In the lattice basis,
$$\theta=(C(G_2), I, C{-1}(G_2)),\quad \omega=(I, C(G_2),
C{-1}(G_2))$$
so that
$$\eqalign{&\theta e_1=-e_1-e_2,\cr &\theta e_2=3e_1+2e_2,}\qquad
\eqalign{&\theta e_3=e_3,\cr &\theta e_4=e_4}\qquad\eqalign{&\theta
e_5=2e_5+e_6,\cr &\theta e_6=-3e_5-e_6}$$
and
$$\eqalign{&\omega e_1=e_1,\cr &\omega e_2=e_2,}\qquad\eqalign{&
\omega e_3=-e_3-e_4,\cr &\omega e_4=3e_3+2e_4,}\qquad
\eqalign{&\omega e_5=2e_5+e_6,\cr&\omega e_6=-3e_5-e_6.}$$
The independent deformation parameters (moduli) are
$R_1$, $R_3$ and $R_5$, where
$$\vert e_2\vert=\sqrt 3\vert e_1\vert\equiv\sqrt 3 R_1,\quad \vert
e_4\vert=\sqrt 3\vert e_3\vert\equiv\sqrt 3 R_3,\quad
\vert e_6\vert=\sqrt 3\vert e_5\vert\equiv\sqrt 3 R_5$$
also, $$\cos{\theta_{12}}=\cos{\theta_{34}}=\cos{\theta_{56}}=-{\sqrt
3\over\displaystyle  2}$$
with all other $\cos{\theta_{ij}}$ zero.
A suitable choice of orthonormal basis $\tilde
e_1,...,\tilde e_6$ is given by
$$\eqalign{&e_1=R_1\tilde e_1,\cr &e_2=R_1(-{3\over\displaystyle
2}\tilde e_1-{\sqrt 3\over\displaystyle 2}\tilde
e_2),}\quad\eqalign{&e_3=R_3\tilde e_3,\cr
&e_4=R_3(-{3\over\displaystyle  2}\tilde e_3-{\sqrt
3\over\displaystyle 2}\tilde e_4),}\quad\eqalign{&e_5=R_5\tilde
e_5,\cr &e_6=R_5(-{3\over\displaystyle  2}\tilde e_5-{\sqrt
3\over\displaystyle 2}\tilde e_6).}$$
The fixed points or tori and associated space group elements are
given with the conventions of table 2 by
\vfill\eject
\begintable
Twisted sector |Fixed point
or torus |$l$\cr$\theta$  |$a_3e_3+a_4e_4
$ |$0$\cr
$\theta2$|$-{\displaystyle n_1\over\displaystyle
3}e_2+{\displaystyle n_5\over\displaystyle  3}e_6+a_3e_3+ a_4
e_4$|$n_1e_1+n_5e_5$\cr
$\theta3$|${\displaystyle n_1\over\displaystyle
2}e_1+{\displaystyle n_2\over\displaystyle  2}e_2+{\displaystyle
n_5\over\displaystyle  2}e_5+{\displaystyle n_6\over\displaystyle
2}e_6+a_3e_3+ a_4e_4$|$n_1e_1+n_2e_2+n_5e_5+n_6e_6$ \cr
$\theta4$ |$-{\displaystyle n_1\over\displaystyle
3}e_2+{\displaystyle n_5\over\displaystyle
3}e_6+a_3e_3+a_4e_4$|$-n_1e_1-n_5e_5$\cr
$\theta5$ |$a_3e_3+
a_4e_4$ |0\cr$\omega$ |$a_1e_1+a_2e_2$ |0\cr
$\omega2$ |$-{\displaystyle n_3\over\displaystyle
3}e_4+{\displaystyle n_5\over\displaystyle
3}e_6+a_1e_1+a_2e_2$|$n_3e_3+n_5e_5$\cr
$\omega3$|${\displaystyle n_3\over\displaystyle
2}e_3+{\displaystyle n_4\over\displaystyle 2}e_4+{\displaystyle
n_5\over\displaystyle  2}e_5+{\displaystyle n_6\over\displaystyle
2}e_6+a_1e_1+a_2 e_2$|$n_3e_3+n_4e_4+n_5 e_5+n_6e_6$\cr
$\omega4$|$-{\displaystyle n_3\over\displaystyle
3}e_4+{\displaystyle n_5\over\displaystyle 3}e_6+a_1e_1+a_2
e_2$|$-n_3e_3-n_5e_5$\cr $\omega5$|$a_1e_1+a_2e_2$
|0\cr$\theta\omega$ |${\displaystyle n_5\over\displaystyle 3}e_6$
|$n_5e_5$\cr$\theta\omega2$ | $-{\displaystyle n_3\over\displaystyle
3}e_4+{\displaystyle n_5\over\displaystyle 2}e_5+{\displaystyle
n_6\over\displaystyle 2}e_6$ |
$n_3e_3+n_5e_5+n_6e_6$\cr$\theta\omega3$ |${\displaystyle
n_3\over\displaystyle 2}e_3+ {\displaystyle n_4\over\displaystyle
2}e_4+{\displaystyle n_5\over\displaystyle 3}e_6$
|$n_3e_3+n_4e_4-n_5e_5$ \cr$\theta\omega4$ |$ -{\displaystyle
n_3\over\displaystyle 3}e_4$ |$-n_3e_3 $\cr$\theta\omega5$
|$a_5e_5+a_6e_6$
 |$0$\cr$\theta2\omega$ |$-{\displaystyle n_1\over\displaystyle
3}e_2+ {\displaystyle n_5\over\displaystyle 2}e_5+{\displaystyle
n_6\over\displaystyle 2}e_6$ |$n_1e_1+n_5e_5+n_6e_6$\cr$
\theta2\omega2$ |$-{\displaystyle n_1\over\displaystyle
3}e_2-{\displaystyle n_3\over\displaystyle 3}e_4+{\displaystyle
n_5\over\displaystyle 3}e_6 $
|$n_1e_1+n_3e_3-n_5e_5$\cr$\theta2\omega3$ |$ -{\displaystyle
n_1\over\displaystyle 3}e_2+{\displaystyle n_3\over\displaystyle
2}e_3+{\displaystyle n_4\over\displaystyle 2}e_4$ |
$n_1e_1+n_3e_3+n_4e_4$ \cr$\theta2\omega4$| $-{\displaystyle
n_1\over\displaystyle 3}e_2-{\displaystyle n_3\over\displaystyle
3}e_4+a_5e_5+a_6e_6$| $n_1e_1-n_3e_3$\cr $\theta2\omega5$
|$-{\displaystyle n_1\over\displaystyle 3}e_2$ |$n_1e_1$\cr
$\theta3\omega$ |${\displaystyle n_1\over\displaystyle
2}e_1+{\displaystyle n_2\over\displaystyle 2}e_2+{\displaystyle
n_5\over\displaystyle 3}e_6$
 |$n_1e_1+n_2e_2-n_5e_5$\cr$\theta3\omega2$ |${\displaystyle
n_1\over\displaystyle 2}e_1+ {\displaystyle n_2\over\displaystyle
2}e_2-{\displaystyle n_3\over\displaystyle 3}e_4$
|$n_1e_1+n_2e_2+n_3e_3$\cr$ \theta3\omega3$|${\displaystyle
n_1\over\displaystyle 2}e_1+ {\displaystyle n_2\over\displaystyle
2}e_2+ {\displaystyle n_3\over\displaystyle 2}e_3+{\displaystyle
n_4\over\displaystyle 2}e_4+a_5e_5+a_6e_6$
|$n_1e_1+n_2e_2+n_3e_3+n_4e_4$ \endtable Continued on next page
\vfill\eject
\begintable
$\theta3\omega4$
|$ {\displaystyle n_1\over\displaystyle 2}e_1+{\displaystyle
n_2\over\displaystyle 2}e_2-{\displaystyle n_3\over\displaystyle
3}e_4$| $n_1e_1+n_2e_2-n_3e_3$ \cr $\theta3\omega5$|
${\displaystyle n_1\over\displaystyle 2}e_1+{\displaystyle
n_2\over\displaystyle 2}e_2+{\displaystyle n_5\over\displaystyle
2}e_5+{\displaystyle n_6\over\displaystyle 2}e_6$|
$n_1e_1+n_2e_2+n_5e_5+n_6e_6$\cr $\theta4\omega$|$-{\displaystyle
n_1\over\displaystyle 3}e_2$|$-n_1e_1$\cr
$\theta4\omega2$|$-{\displaystyle n_1\over\displaystyle
3}e_2-{\displaystyle n_3\over\displaystyle 3}e_4+a_5e_5+a_6e_6$
 |$-n_1e_1+n_3e_3$\cr
$\theta4\omega3$
|$-{\displaystyle n_1\over\displaystyle 3}e_2+ {\displaystyle
n_3\over\displaystyle 2}e_3+{\displaystyle n_4\over\displaystyle
2}e_4$ |$-n_1e_1+n_3e_3+n_4e_4$\cr
$\theta4\omega4$|$-{\displaystyle n_1\over\displaystyle 3}e_2-
{\displaystyle n_3\over\displaystyle 3}e_4+{\displaystyle
n_5\over\displaystyle 3}e_6$|$-n_1e_1-n_3e_3+n_5e_5$
\cr$\theta4\omega5$|$ -{\displaystyle n_1\over\displaystyle
3}e_2+{\displaystyle n_5\over\displaystyle 2}e_5+{\displaystyle
n_6\over\displaystyle 2}e_6$| $-n_1e_1+n_5e_5+n_6e_6$
\cr$\theta5\omega$|$a_5e_5+a_6e_6$|$0$ \cr
$\theta5\omega2$|$-{\displaystyle n_3\over\displaystyle
3}e_4$|$n_3e_3$\cr $\theta5\omega3$|$ {\displaystyle
n_3\over\displaystyle 2}e_3+{\displaystyle n_4\over\displaystyle
2}e_4+{\displaystyle n_5\over\displaystyle 3}e_6$|
$n_3e_3+n_4e_4+n_5e_5$ \cr$\theta5\omega4$| $-{\displaystyle
n_3\over\displaystyle 3}e_4+{\displaystyle n_5\over\displaystyle
2}e_5+{\displaystyle n_6\over\displaystyle 2}e_6$|
$-n_3e_3+n_5e_5+n_6e_6$\cr $\theta5\omega5$|${\displaystyle
n_5\over\displaystyle 3}e_6$|$-n_5e_5$ \endtable \vfill\eject
The action of $\theta$ and $\omega$ on the fixed points (or tori)
is given by
\vskip 0.5cm
\begintable
twisted |Action of $\theta$
 |Action of $\omega$\nr sector| | \cr
$\theta,\theta5$
 |$I$ |$I$
\cr$\theta2,\theta4$ |$(n_1,n_5)\rightarrow (-n_1,-n_5)$ |
$n_5\rightarrow -n_5$
\cr$\theta3$
|$(n_1,n_2,n_5,n_6)\rightarrow (n_1+n_2,n_1,n_6,n_5+n_6)$|
$(n_5,n_6)\rightarrow (n_6,n_5+n_6)$
\cr$\omega,\omega5$ |$I$|$I$
\cr$\omega2,\omega4$ |$n_5\rightarrow-n_5$
 |$(n_3,n_5)\rightarrow
(-n_3,-n_5)$
\cr$\omega3$ |$(n_5,n_6)\rightarrow
(n_6,n_5+n_6)$ |$(n_3,n_4,n_5,n_6)\rightarrow
(n_3+n_4,n_3,n_6,n_5+n_6)$\cr
$\theta\omega,\theta5\omega5$
 |$n_5\rightarrow
-n_5$ |$n_5\rightarrow -n_5$
\cr$\theta\omega2,\theta5\omega4$ |$(n_5,n_6)\rightarrow
(n_6,n_5+n_6)$ |$(n_3,n_5,n_6)\rightarrow (-n_3,n_6,n_5+n_6)$
\cr$\theta\omega3,\theta5\omega3$ |$n_5\rightarrow -n_5$
|$(n_3,n_4,n_5)\rightarrow (n_3+n_4,n_3,-n_5)$
\cr$\theta\omega4,\theta5\omega2$ |$I$ |$n_3\rightarrow
-n_3$
\cr$\theta\omega5,\theta5\omega$ |$I$ |$I$
\cr$\theta2\omega,\theta4\omega5$ |$(n_1,n_5,n_6)\rightarrow
(-n_1,n_6,n_5+n_6)$|$(n_5,n_6)\rightarrow
\rightarrow (n_6,n_5+n_6)$
\cr$\theta2\omega2,\theta4\omega4$ |$(n_1,n_5)\rightarrow
(-n_1,-n_5)$ |$(n_3,n_5)\rightarrow
(-n_3,-n_5)$
\cr$\theta2\omega3,\theta4\omega3$ |$n_1\rightarrow
-n_1$ |$(n_3,n_4)\rightarrow
(n_3+n_4,n_3)$
\cr$\theta2\omega4,\theta4\omega2$ |$n_1\rightarrow
-n_1$ |$n_3\rightarrow
-n_3$
\cr$\theta2\omega5,\theta4\omega$ |$n_1\rightarrow
-n_1$ |$I$
\cr$\theta3\omega,\theta3\omega5$ |$
(n_1,n_2,n_5)\rightarrow
(n_1+n_2,n_1,-n_5)$|$n_5\rightarrow
-n_5$
\cr$\theta3\omega2,\theta3\omega4$ |$(n_1,n_2)\rightarrow
(n_1+n_2,n_1)$ |$n_3\rightarrow
-n_3$
\cr$\theta3\omega3$|$
(n_1,n_2)\rightarrow
(n_1+n_2,n_1)$ |$(n_3,n_4)\rightarrow
(n_3+n_4,n_3)$
\endtable
The moduli dependent suppression factors for the non-zero Yukawa
couplings amongst twisted sectors containing massless states are
given in the conventions of table 4 by
\vfill\eject
\begintable
Yukawa
Coupling|$\lambda_1$|$\lambda_2$|$\lambda_3$\cr
$T_{01}T_{14}T_{51}$|0|1 for $n_3(T_{14})\not=0$|0\cr
$T_{01}T_{15}T_{50}$|0|0|0\cr
$T_{01}T_{23}T_{42}$|0|
$\mu(n_3, p_3, p_4)$|0\cr
$T_{01}T_{24}T_{41}$|0|1 for $n_3(T_{24})\not=0$|0\cr
$T_{01}T_{32}T_{33}$|0|$\mu(n_3, p_3, p_4)$|0\cr
$T_{02}T_{13}T_{51}$|0|$\mu(n_3, p_3, p_4)$|0\cr
$T_{02}T_{14}T_{50}$|0|0|1 for $n_5(T_{02})\not=0$\cr
$T_{02}T_{22}T_{42}$|0|2 for $m_3(T_{02})-n_3(T_{22})\not=0$|0\cr
$T_{02}T_{23}T_{41}$|0|$\mu(n_3, p_3, p_4)$|1 for $n_5(T_{02})\not=0$\cr
$T_{02}T_{24}T_{40}$|0|0|0\cr
$T_{02}T_{31}T_{33}$|0|$\mu(n_3, p_3, p_4)$|0\cr
$T_{02}T_{32}T_{32}$|0|2 for $m_3(T_{02})-n_3(T_{32})\not=0$|
1 for $n_5(T_{02})\not=0$\cr
$T_{03}T_{12}T_{51}$|0|$\mu(n_3, p_3, p_4)$|0\cr
$T_{03}T_{13}T_{50}$|0|0|$\mu(n_5, p_5, p_6)$\cr
$T_{03}T_{21}T_{42}$|0|$\mu(n_3, p_3, p_4)$|0\cr
$T_{03}T_{22}T_{41}$|0|$\mu(n_3, p_3, p_4)$|
$\mu(n_5, p_5, p_6)$\cr
$T_{03}T_{23}T_{40}$|0|0|
$\mu(n_5, p_5, p_6)$\cr
$T_{03}T_{30}T_{33}$|0|0|
0\cr
$T_{03}T_{31}T_{32}$|0|$\mu(n_3, p_3, p_4)$|$\mu(n_5, p_5, p_6)$\cr
$T_{04}T_{11}T_{51}$|0|1 for $n_3(T_{04})\not=0$|0\cr
$T_{04}T_{12}T_{50}$|0|0|$\mu(n_5, p_5, p_6)$\cr
$T_{04}T_{20}T_{42}$|0|0|0\cr
$T_{04}T_{21}T_{41}$|0|1 for
$n_3(T_{04})\not=0$|$\mu(n_5, p_5, p_6)$\cr
$T_{04}T_{22}T_{40}$|0|0|2 for
$n_5(T_{04})-n_5(T_{22})\not=0$
\endtable
Continued on next page
\vfill\eject
\begintable
$T_{04}T_{30}T_{32}$|0|0|$\mu(n_5, p_5, p_6)$\cr
$T_{05}T_{10}T_{51}$|0|0|0\cr
$T_{05}T_{11}T_{50}$|0|0|1 for
$n_5(T_{11})\not=0$\cr
$T_{05}T_{20}T_{41}$|0|0|1 for
$n_5(T_{20})\not=0$\cr
$T_{05}T_{21}T_{40}$|0|0|$\mu(n_5, p_5, p_6)$\cr
$T_{05}T_{30}T_{31}$|0|0|$\mu(n_5, p_5, p_6)$\cr
$T_{10}T_{14}T_{42}$|1 for
$n_1(T_{42})\not=0$|0|0\cr
$T_{10}T_{15}T_{41}$|1 for
$n_1(T_{41})\not=0$|0|0\cr
$T_{10}T_{23}T_{33}$|$\mu(n_1, p_1, p_2)$|0|0\cr
$T_{10}T_{24}T_{32}$|$\mu(n_1, p_1, p_2)$|0|0\cr
$T_{11}T_{13}T_{42}$|1 for
$n_1(T_{42})\not=0$|$\mu(n_3, p_3, p_4)$|0\cr
$T_{11}T_{14}T_{41}$|1 for
$n_1(T_{41})\not=0$|1 for
$n_3(T_{14})\not=0$|1 for
$n_5(T_{11})\not=0$\cr
$T_{11}T_{15}T_{40}$|1 for
$n_1(T_{40})\not=0$|0|0\cr
$T_{11}T_{22}T_{33}$|$\mu(n_1, p_1, p_2)$|$\mu(n_3, p_3, p_4)$|0\cr
$T_{11}T_{23}T_{32}$|$\mu(n_1, p_1, p_2)$|$\mu(n_3, p_3,
p_4)$|1 for $n_5(T_{11})\not=0$\cr
$T_{11}T_{24}T_{31}$|$\mu(n_1, p_1, p_2)$|1 for
$n_3(T_{24})\not=0$|0\cr
$T_{12}T_{12}T_{42}$|1 for
$n_1(T_{42})\not=0$|2 for $m_3(T_{12})-n_3(T_{12})\not=0$|0\cr
$T_{12}T_{13}T_{41}$|1 for $n_1(T_{41})\not=0$|$\mu(n_3, p_3, p_4)$|
$\mu(n_5, p_5, p_6)$\cr
$T_{12}T_{14}T_{40}$|1 for $n_1(T_{40})\not=0$|0|$\mu(n_5, p_5,
p_6)$\cr
$T_{12}T_{21}T_{33}$|$\mu(n_1, p_1, p_2)$|$\mu(n_3, p_3,
p_4)$|0\cr
$T_{12}T_{22}T_{32}$|$\mu(n_1, p_2, p_2)$|2 for
$m_3(T_{12})-n_3(T_{22})\not=0$|$\mu(n_5, p_5, p_6)$\cr
$T_{12}T_{23}T_{31}$|$\mu(n_1, p_1, p_2)$|$\mu(n_3, p_3,
p_4)$|$\mu(n_5, p_5, p_6)$\cr
$T_{12}T_{24}T_{30}$|$\mu(n_1, p_1, p_2)$|0|0\cr
$T_{13}T_{13}T_{40}$|1 for $n_1(T_{40})\not=0$|0|2 for
$m_5(T_{13})-n_5(T_{13})\not=0$\cr
$T_{13}T_{20}T_{33}$|$\mu(n_1, p_1, p_2)$|0|0
\endtable
Continued on next page
\vfill\eject
\begintable
$T_{13}T_{21}T_{32}$|$\mu(n_1, p_1, p_2)$|$\mu(n_3, p_3, p_4)$|$
\mu(n_5, p_5, p_6)$\cr
$T_{13}T_{22}T_{31}$|$\mu(n_1, p_1, p_2)$|$\mu(n_3, p_3, p_4)$|2 for
$m_5(T_{22})-n_5(T_{13})\not=0$\cr
$T_{13}T_{23}T_{30}$|$\mu(n_1, p_1, p_2)$|0|$\mu(n_5, p_5, p_6)$\cr
$T_{14}T_{20}T_{32}$|$\mu(n_1, p_1, p_2)$|0|1 for $n_5(T_{20})\not=0$
\cr
$T_{14}T_{21}T_{31}$|$\mu(n_1, p_1, p_2)$|1 for
$n_3(T_{14})\not=0$|$\mu(n_5, p_5, p_6)$\cr
$T_{14}T_{22}T_{30}$|$\mu(n_1, p_1, p_2)$|0|$\mu(n_5, p_5, p_6)$\cr
$T_{15}T_{20}T_{31}$|$\mu(n_1, p_1, p_2)$|0|0\cr
$T_{15}T_{21}T_{30}$|$\mu(n_1, p_1, p_2)$|0|0\cr
$T_{20}T_{22}T_{24}$|2 for $m_1(T_{20})-n_1(T_{22})\not=0$|0|0\cr
$T_{20}T_{23}T_{23}$|2 for $m_1(T_{20})-n_1(T_{23})\not=0$|1 for
$n_3(T_{20})\not=0$|0\cr
$T_{21}T_{21}T_{24}$|2 for
$m_1(T_{21})-n_1(T_{21})\not=0$|1 for
$n_3(T_{24})\not=0$|0\cr
$T_{21}T_{22}T_{23}$|2 for
$m_1(T_{21})-n_1(T_{22})\not=0$|$\mu(n_3, p_3, p_4)$|
$\mu(n_5, p_5, p_6)$\cr
$T_{22}T_{22}T_{22}$|2 for $m_1(T_{22})-n_1(T_{22})\not=0$| 2 for
$m_3(T_{22})-n_3(T_{22})\not=0$| 2 for
$m_5(T_{22})-n_5(T_{22})\not=0$
\endtable
\bye